\documentclass{article}

\usepackage{arxiv}

\usepackage[utf8]{inputenc} 
\usepackage[T1]{fontenc}    
\usepackage{hyperref}       
\usepackage{url}            
\usepackage{booktabs}       
\usepackage{amsfonts}       
\usepackage{nicefrac}       
\usepackage{microtype}      
\usepackage{doi}
\usepackage{subcaption}     
\usepackage{bbm}
\usepackage{multirow}

\usepackage{amssymb}
\usepackage{amsmath}
\usepackage{mathtools} 
\usepackage{xcolor}
\usepackage{media9} 
\usepackage{graphicx, animate}
\usepackage{units} 
\usepackage{esvect}
\usepackage{comment}
\usepackage{cite} 
\usepackage{float}
\usepackage{bm}

\usepackage[labelfont={bf,sf},font={small}, labelsep=space]{caption}
\usepackage{algorithm}
\usepackage{algpseudocode}

\usepackage{listings}

\definecolor{darkred}{rgb}{0.6, 0, 0}
\definecolor{softgreen}{rgb}{0.2, 0.6, 0.2}

\usepackage{todonotes}

\newcommand{\divergence}{\mathop\mathrm{div}}

\newcommand{\sY}{\hspace{-0.05cm} \prescript{S}{}{Y}}

\title{Mobility-informed coupling of ABM, PDE, and ODE models for pandemic simulation in Germany} 

\date{} 					

\author{ 
{
Kristina Kehrer}
\\
	Department of Visual and Data-Centric Computing\\
	Zuse Institute Berlin\\
        Takustraße 7 \\
	Berlin, 14195 \\
        Germany \\
	\texttt{kehrer@zib.de} \\
	\And
 {
Tim O. F. Conrad} \\
	Department of Visual and Data-Centric Computing\\
	Zuse Institute Berlin\\
        Takustraße 7 \\
	Berlin, 14195 \\
        Germany \\
	\texttt{conrad@zib.de} \\
}

\renewcommand{\headeright}{Paper}

\allowdisplaybreaks 

\begin{document}
\maketitle
\begin{abstract} 
Simulating epidemic spread across an entire country requires balancing fine-grained realism with computational feasibility. We address this trade-off with a multiscale, hybrid modeling framework for simulating the spread of COVID-19 across Germany. The spatial domain is decomposed into multiple regions, each of which can be represented either by a high-resolution agent-based model (ABM) incorporating mobility data from mobile phones or by a faster, less detailed deterministic model based on partial differential equations (PDEs) or ordinary differential equations (ODEs). Mobility between regions is incorporated through data-driven jump processes, enabling individuals to be transferred between model domains. Building on earlier studies on pairwise ABM-ODE, ABM-PDE, and PDE-ODE coupling strategies, we develop a unified framework that combines all three model classes within a single simulation environment. To demonstrate the framework's utility, we systematically compare ABM, PDE, and ODE representations of Berlin embedded in a nationwide simulation of Germany, investigate regional travel restrictions, and evaluate the Zero-COVID and No-COVID strategies. 
The results indicate that model resolution can be reduced in sufficiently homogeneous regions without substantially altering epidemic dynamics. Furthermore, the intervention experiments reveal that mobility restrictions can lead to non-intuitive outcomes that reflect the nonlinear structure of the epidemic dynamics, including cases in which regional border closures increase infection numbers both locally and nationally. These effects are observed even between non-adjacent regions, driven by long-distance mobility coupling rather than geographic proximity,  illustrating how emergent, system-wide dynamics arise from local mobility restrictions. We quantify computational performance in terms of runtime savings and validate the framework against real-world infection data. The results show that the hybrid framework substantially reduces computational cost without sacrificing predictive accuracy, offering a practical tool for evaluating regional mobility restrictions and public health interventions at national scale.
\end{abstract}

\keywords{Partial differential equation \and ordinary differential equation \and epidemic modeling \and spatial modeling \and coupling approach \and diffusion \and agent-based model \and landscape \and mobility}

\section{Introduction}

Timely and effective public health countermeasures depend on accurate, computationally efficient simulation of infection dynamics. In a highly interconnected country such as Germany, mobility and interactions across regional borders shape the spread of disease and must therefore be incorporated into epidemiological models to support interventions that reduce mortality and prevent critical healthcare overload. 

In this paper, we introduce a multiscale, hybrid modeling framework that combines agent-based models (ABMs) with partial differential equation (PDE) and ordinary differential equation (ODE) models to balance accuracy and computational scalability. The agent-based component captures fine-grained infection dynamics and mobility patterns derived from mobile phone data, while the differential equation models provide a coarser but faster representation of regions where purely agent-based simulation is computationally prohibitive. Data-driven jump processes determine the number of individuals transferred between model domains, coupling regions into a mobility network that connects the individually simulated federal states, ensuring a consistent integration of dynamics across the entire system regardless of which model types are coupled. To enhance realism, we further incorporate temporal changes in human behavior through reductions in activities performed outside the home, which we refer to throughout this work as activity reductions.

This study builds directly on our own prior work, which addressed these couplings only pairwise: ABM-ODE~\cite{bostanci2025integrating}, ABM-PDE~\cite{Kehrer2026HybridAbmPde}, and PDE-ODE~\cite{Kehrer2025Hybrid}. Here, we unify all three model types within a single hybrid framework and apply the resulting approach to Germany as a whole, enabling large-scale simulations that retain local detail where necessary while keeping computational cost manageable. 

ABMs have previously been employed at the national scale in Germany to study complex socio-economic processes, demonstrating their ability to represent heterogeneous agents, social interactions, and decision-making mechanisms \cite{Krebs2017GreenABM,Alyousef2017PVBattery}. 
In the context of infectious disease modeling, similar large-scale frameworks have been developed to simulate entire populations, incorporating demographic data, mobility networks, and policy interventions \cite{Bullock2021JuneABMEngland,Heger2025JuneABMGermanyReviewed}. 
However, these approaches rely on a single model type throughout, so their computational cost scales directly with the size of the population represented at agent-based resolution. Hybrid approaches that couple detailed and reduced models have instead been explored at regional scales, demonstrating their potential to capture spatial heterogeneity while reducing computational cost\cite{Bicker2025Hybrid}, but have not yet been applied at national scale. Our framework closes this gap by combining mobility-driven ABMs with reduced PDE and ODE models across an entire country. 

The present work builds on a streamlined C++ implementation of EpiSim~\cite{Muller2021ABM,GithubEpiSim}, an established agent-based simulation framework that incorporates mobility patterns derived from mobile phone data to generate realistic daily routines. To reduce preprocessing complexity and computational demands, simulations are conducted on a representative subset corresponding to 25\% of the population, enabling efficient large-scale analysis while preserving population-level epidemic dynamics. 

The software developed in this work, together with a subset of input data, including 
PDE grid information, is publicly available via Zenodo~\cite{kehrer2026AbmPdeOdeCode}. 
Additional datasets, in particular processed mobility data such as events, facility information, initial agent distributions, landscape representations, 
are not publicly distributed but can be provided by the authors upon reasonable request.

\section{Models and Reductions} \label{sec:Models}
In the following, we introduce the three model types employed in the hybrid framework and describe how they are initialized and coupled across regions. Germany's 16 federal states are each assigned to one of these model types (ABM, PDE, or ODE model) based on their computational characteristics. In federal states with sufficiently large populations, aggregated models such as PDEs or ODEs provide a suitable approximation of the underlying dynamics, as stochastic fluctuations average out. If mixing is sufficiently homogeneous, ODE models are appropriate, whereas PDE models can be used when spatial heterogeneity needs to be resolved. In contrast, ABMs can be employed where individual-level heterogeneity or detailed mobility patterns are essential for accurately capturing the dynamics. By choosing an appropriate model type for each federal state, the hybrid framework can reduce computational cost while preserving accuracy.

\begin{figure}[h!]
    \centering
    \includegraphics[width=0.4\textwidth]{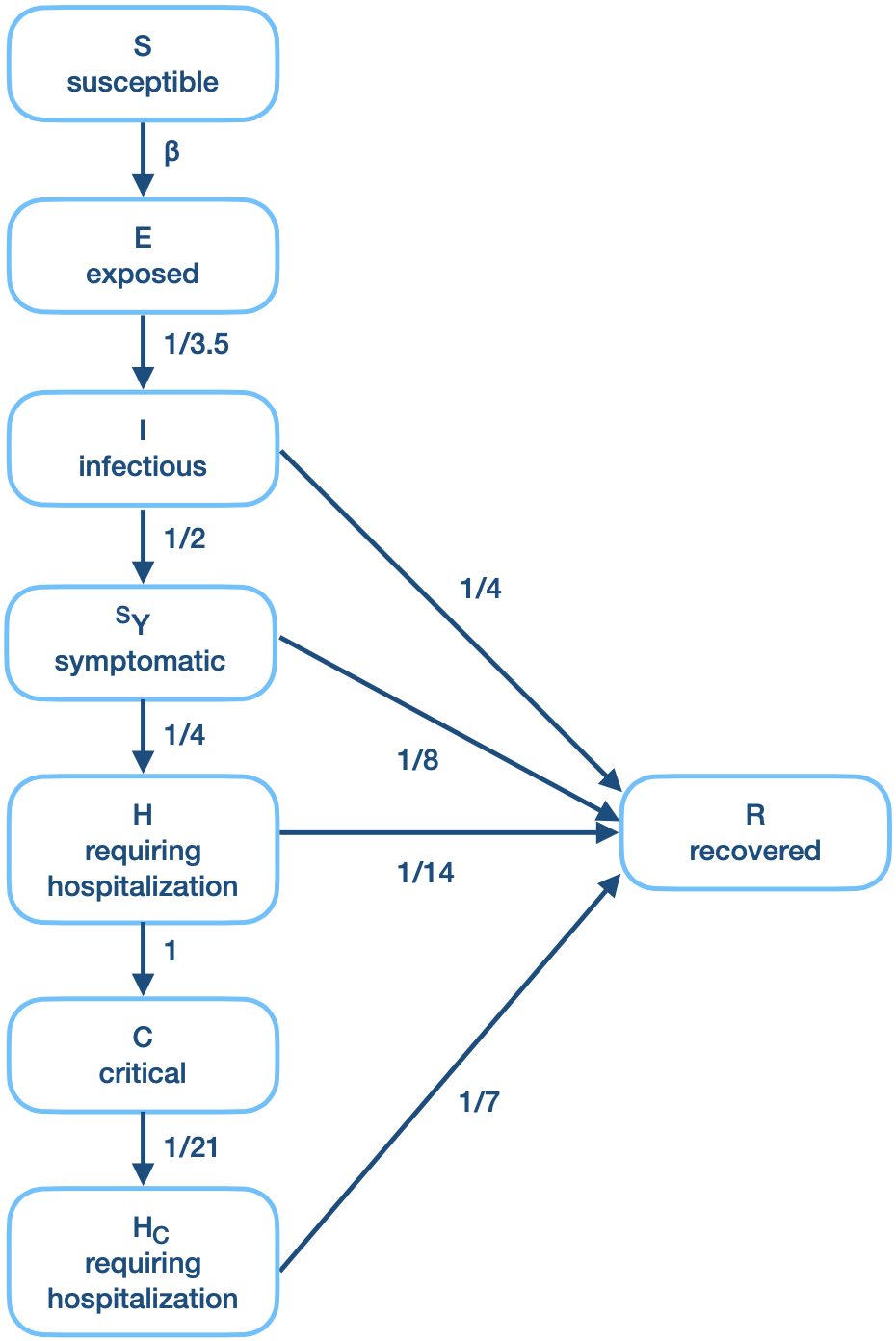}
    \caption{Structure of the SEIYHCR model with seven health states and ten transition rules, capturing disease progression from \textbf{S}usceptble to \textbf{R}ecovered via intermediate states.}
    \label{fig:compartments_flow}
\end{figure}

All three model types incorporate the same epidemiological health states. The health statuses in our model are defined as follows:
\begin{itemize}
    \item susceptible $\left(S\right)$
    \item exposed $\left(E\right)$ - not symptomatic, not infectious
    \item infectious $\left(I\right)$ - not symptomatic, infectious
    \item symptomatic $\left(\,\sY\right)$ - symptomatic, infectious
    \item requiring hospitalization (after being symptomatic) $\left(H\right)$ 
    \item critical $\left(C\right)$ 
    \item requiring hospitalization (after being in critical state) $\left(H_C\right)$
    \item recovered $\left(R\right)$.
\end{itemize}
We do not consider deaths or births. Furthermore, movements across the German national border are not represented in the underlying mobility data, such that the total population across all models remains constant over time. 
The chosen health states (compartments) and transition rates between compartments are displayed in Fig~\ref{fig:compartments_flow}. The transition rates are chosen values from the literature except for the infection rate $\beta$, whose selection is discussed separately in Section~\ref{sec:ParameterOptimization}.
\noindent

All model domains are dynamically coupled at each time step: individuals transitioning between regions are transferred between the corresponding domains. Transitions from the ABM to non-ABM (PDE or ODE) domains are represented either as spatial density contributions in the PDE or as aggregated compartment counts in the ODE. Conversely, individuals from PDE or ODE domains can be mapped back to the ABM. Transitions between non-ABM domains are also supported. Direct transitions between ABM domains are not considered, since all ABM regions are treated as parts of a single global ABM representation. 
The decision of when and how many individuals transition depends on the evaluated mobility data. This exchange preserves the total population across model interfaces and reflects observed mobility patterns.

The remainder of this section is structured as follows:
\begin{itemize}
    \item Part~1 introduces an event-based ABM, based on mobile phone data (Section~\ref{sec:ABM}). 
    \item Part~2 presents the derivation of the PDE-based compartment model (Section~\ref{sec:PDEModelDerivation}) using a motion-based ABM, which is related to the event-based ABM.
    \item Part~3 presents the derivation of the ODE-based compartment model (Section~\ref{sec:ODEModelDerivation}) using the PDE model.
    \item The initialization -- or setting of initial conditions -- of all model components is described in Section~\ref{sec:InitialValues}.
    \item Finally, Section~\ref{sec:Coupling} explains the coupling mechanism that integrates all parts into a single hybrid simulation framework.
\end{itemize}

The modeling framework builds on our previous work~\cite{Kehrer2026HybridAbmPde}. The essential components are summarized here for completeness, while modifications are described throughout this Section~\ref{sec:Models} and the subsequent Section~\ref{sec:Implementation}. 

\subsection{Part 1: ABM Based on Mobile Phone Trajectories} \label{sec:ABM}

In our paper, the ABM used in our coupling approach is based on high-resolution mobile phone data. The given data are event-based. Using these data, we have reconstructed individual trajectories. To be precise, the model represents a streamlined version of EpiSim~\cite{Muller2021ABM,GithubEpiSim}. 
Further details on the underlying ABM 
can be found in~\cite{Kehrer2026HybridAbmPde}.

Each agent is either located inside a facility -- such as a home, workplace, school or leisure facility -- or commuting between facilities. 
Within facilities, agents can only interact with others inside the same facility and category. This means that infection transmission can occur only if an infectious ($I$ or $\sY$) and a susceptible ($S$) agent are present in the same location and facility category. 
While commuting, agents are assumed to be isolated and cannot interact and therefore cannot transmit infections. 

Health status updates occur at every time step, except for susceptible agents for which the health status is updated only upon leaving a facility. 
Consistent with the update mechanism of the underlying EpiSim framework, this modeling choice for susceptible agents ensures that the infection process reflects accumulated exposure during the stay. 

Stochastic state transitions are evaluated at discrete time points. Further details on the disease progression dynamics are provided in~\cite{Helfmann2021Interacting,Kehrer2026HybridAbmPde} and in the code (see file \texttt{ABM.cpp}).

\subsection{Part 2: PDE-Based Compartment Model} \label{sec:PDEModelDerivation}

Next, we approximate an ABM with a PDE model. As in our previous paper~\cite{Kehrer2026HybridAbmPde}, we first formulate a motion-based ABM on a given domain -- here Germany -- in such a way that it can be approximated by a system of stochastic PDEs and, by removing the noise, deterministic PDEs. For this purpose, we use mobile phone data to create a landscape that drives 
motion. This results in a PDE system for the entire domain of Germany. Later, we will choose which federal state in Germany will be equipped with which kind of model (see Section~\ref{subsec:ModelTypes}). This means that we may end up with islands in the domain that simulate the infection spread in Germany using PDE (or ODE) systems. 

Actually, we would like to equip each federal state with its own PDE system, even though some of them might share borders, and tune its own infection rates. Hence, we create landscapes for each federal state that will be modeled using a PDE system. By restricting the landscape of Germany to the domain of a federal state, we can formulate PDE systems for each federal state individually. We impose zero Neumann boundary conditions, ensuring that there is no net movement of individuals across the boundary of the PDE domain. Instead, we directly enforce commuting by implementing jumps based on the mobile phone data. This way, we stay close to real-world data and avoid dealing with diffusion through the boundary, which is difficult to 
interpret as a real-world mobility process.

Based on the approach in~\cite{Helfmann2021Interacting}, we can construct a motion-based ABM that can be reduced to a system of stochastic PDEs. The number of equations is equal to the number of compartments (see Fig~\ref{fig:compartments_flow}). By constructing a motion-based ABM related to our event-based 
ABM, we can use the same transition rates for the PDE model~\cite{Kehrer2026HybridAbmPde}. 

To approximate the event-based ABM as closely as possible, the landscape $V: \; \Omega_{Be} \rightarrow \mathbb{R}$ remains the primary parameter to be specified in the motion-based ABM. 
We found that the landscape $V$ can be expressed in a way that incorporates the trajectories over our simulation period:
\begin{align} \label{eq:landscape_definition}
    V(X) = - \biggl(\frac{D}{2}\biggr) \log \bigl( p_{st}(X) \bigr),
\end{align}
where $p_{st}(X)$ is the 
probability distribution of agents at location $X$~\cite{Li2023Numerical}. To approximate the probability distribution of agents at each location, we use mobile phone data. 

\begin{figure}[h!]
    \centering
    \includegraphics[width=0.5\linewidth]{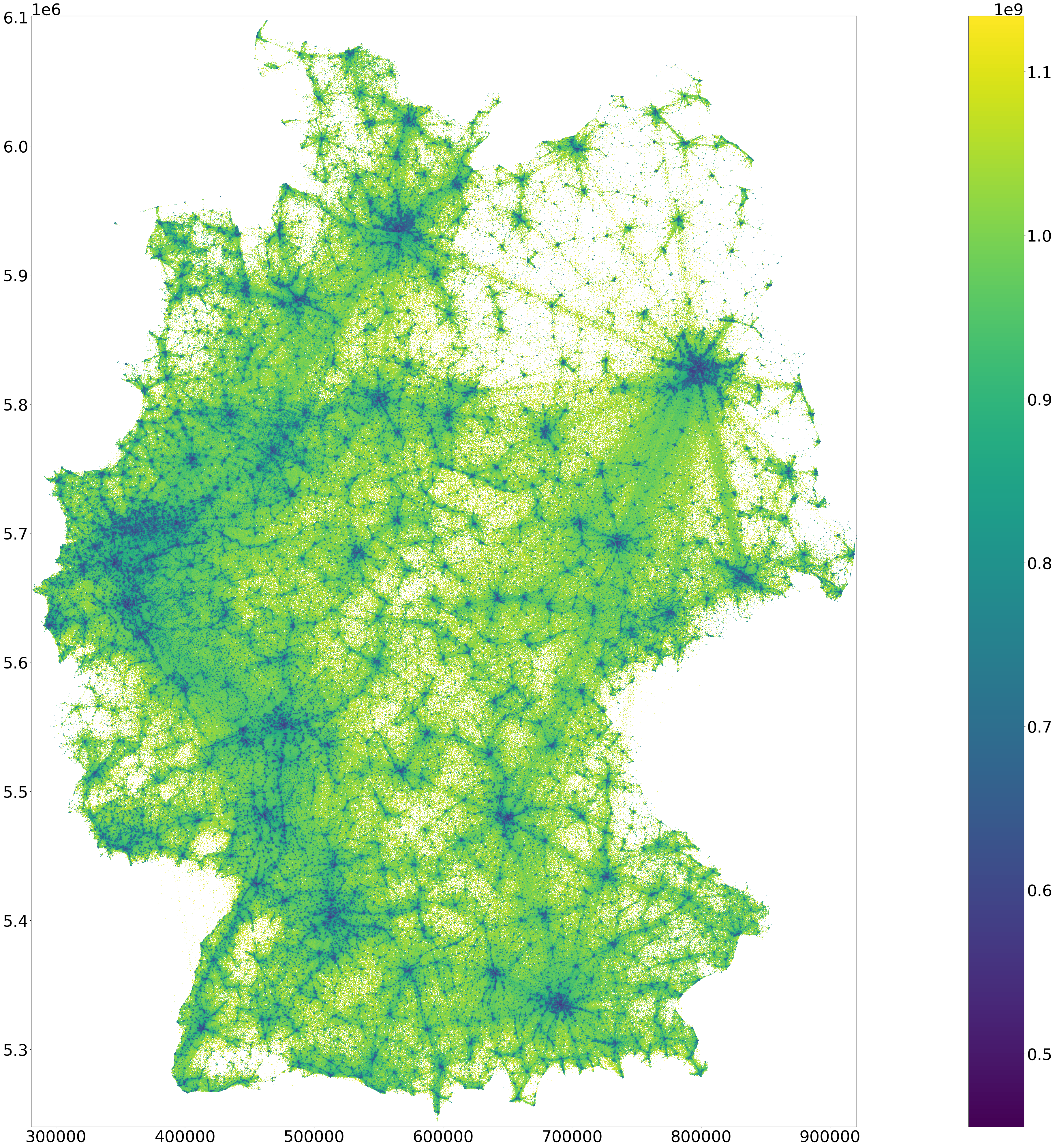}
    \caption{Landscape of Germany}
    \label{fig:landscape}
\end{figure}

First, we generate agent trajectories by determining their location for each hour of the day. 
Next, we aggregate all trajectories for each hour over the entire week to construct a histogram. After normalization, this histogram represents the probability distribution of the agents, which we use to compute the landscape. 
The 
landscape of Germany is shown in Fig~\ref{fig:landscape}.
Since mobile phone data are available for all of Germany, the landscape is initially defined for the entire region, after which we extract the portion corresponding to each federal state modeled using a PDE system.

The resulting PDE for susceptibles in federal state $i$ is then given by 
\begin{align} 
    && \frac{\partial S_i(x,t)}{\partial t} = \; &D \Delta S_i(x,t) + \divergence (\nabla V_i(x) S_i(x,t)) - \beta_i S_i(x,t) \bigl( I_i(t)^{B_r(x)} + \sY_i(t)^{B_r(x)} \bigr).
\end{align}
We have separate PDE systems for each federal state and hence different densities $Y_i\in \{S_i,E_i,I_i,\sY_i,H_i,C_i,H_{C_i},R_i\}$, as well as different infection rates $\beta_i$ and different landscapes $V_i$. For simplicity, we continue without the index $i$: 
\begin{align} 
    && \frac{\partial S(x,t)}{\partial t} = \; &D \Delta S(x,t) + \divergence (\nabla V(x) S(x,t)) - \beta S(x,t) \bigl( I(t)^{B_r(x)} + \sY(t)^{B_r(x)} \bigr). 
\end{align}
Since the spatial mesh used in our simulations is relatively coarse, with grid spacing specified in the input data files, the distance between neighboring grid points is substantial -- on the order of hundreds of meters. In the context of infection spreading, this implies that within a ball $B_r(x)$ of contact radius $r$, it is reasonable to assume that only a single grid point contributes significantly. Hence, we approximate the integral of infectious density over the ball $B_r(x)$ by the infectious density at the center $x$, scaled by the area of the ball: 
\begin{align*}
I(t)^{B_r(x)} + \sY(t)^{B_r(x)} = \int_{B_r(x)} \bigl( I(y,t)+\sY(y,t) \bigr) \; dy \approx \pi r^2 \bigl( I(x,t)+\sY(x,t) \bigr).
\end{align*}
This localized approximation is consistent with the resolution of the spatial discretization and simplifies the infection term.
The resulting system of PDEs is given by 
\begin{align} \label{eq:PDEmodel} 
    && \frac{\partial S(x,t)}{\partial t} \approx \; &D \Delta S(x,t) + \divergence (\nabla V(x) S(x,t)) - \beta \pi r^2 S(x,t) \bigl( I(x,t)+\sY(x,t) \bigr) \notag 
 \\
    && \frac{\partial E(x,t)}{\partial t} \approx \; &D \Delta E(x,t) + \divergence (\nabla V(x) E(x,t)) + \beta \pi r^2 S(x,t) \bigl( I(x,t)+\sY(x,t) \bigr) - \sigma E(x,t) \notag  \\
   && \frac{\partial I(x,t)}{\partial t} = \; &D \Delta I(x,t) + \divergence (\nabla V(x) I(x,t)) + \sigma E(x,t) - \phi_i I(x,t) - \gamma I(x,t) \notag  \\
   && \frac{\partial \sY(x,t)}{\partial t}= \; &D \Delta \sY(x,t)+ \divergence (\nabla V(x) \sY(x,t))+ \gamma I(x,t) - \phi_{sy} \sY(x,t) - \eta \sY(x,t)  \\
   && \frac{\partial H(x,t)}{\partial t} = \; &D \Delta H(x,t) + \divergence (\nabla V(x) H(x,t)) + \eta \sY(x,t)  - \phi_{h} H(x,t) - \kappa H(x,t) \notag  \\
   && \frac{\partial C(x,t)}{\partial t} = \; &D \Delta C(x,t) + \divergence (\nabla V(x) C(x,t)) + \kappa H(x,t)                   - \eta_c C(x,t) \notag  \\
   &&\frac{\partial H_C(x,t)}{\partial t}=\; &D \Delta H_C(x,t)+\divergence (\nabla V(x) H_C(x,t))+ \eta_c C(x,t) - \phi_{hc} H_C(x,t)     \notag          \\
   && \frac{\partial R(x,t)}{\partial t} = \; &D \Delta R(x,t) + \divergence (\nabla V(x) R(x,t)) + \phi_i I(x,t) + \phi_{sy} \sY(x,t) + \phi_{h} H(x,t) + \phi_{hc} H_C(x,t). \notag 
\end{align}
To prepare the PDE system~\eqref{eq:PDEmodel} for numerical simulation using the finite element method, we derive its weak formulation. This involves multiplying each equation by a test function $\varphi$ and integrating over the domain $\Omega_{\text{PDE}}$ of each federal state represented by a reduced non-ABM model. In this way, we obtain the weak formulation of the PDE system for the compartment densities. The weak formulation corresponding to the susceptible density equation is given by
\begin{align*} 
    \int_{\Omega_{\text{PDE}}} \frac{\partial S(x,t)}{\partial t} \varphi \; dx 
    &\approx
    \int_{\Omega_{\text{PDE}}} D \Delta S(x,t) \varphi + \divergence (\nabla V(x) S(x,t)) \varphi - \beta \pi r^2 S(x,t) \bigl( I(x,t)+\sY(x,t) \bigr) \varphi \; dx
    \\ &= \int_{\Omega_{\text{PDE}}} - \nabla \varphi ^T D \nabla S  - \nabla \varphi ^T (S \nabla V) - \beta \pi r^2 S \bigl( I+\sY \bigr) \varphi \; dx \\ &\quad+ \int_{\Gamma_{\text{PDE}}} \varphi  \underbrace{\nu^T (D \nabla S + S \nabla V)}_{=0 \text{ on } \Gamma_{\text{PDE}}} \; do
    \\ &= 
    \int_{\Omega_{\text{PDE}}} - \nabla \varphi ^T D \nabla S  - \nabla \varphi ^T (S \nabla V) - \beta \pi r^2 S \bigl( I+\sY \bigr) \varphi \; dx,
\end{align*}
where $\nu$ denotes the unit outer normal to $\Omega_{\text{PDE}}$. 
Analogous approximations are applied to the weak formulations of the other compartments $E$, $I$, $\sY$, $H$, $C$, $H_C$, and $R$, which follow the same structure. For implementation details, we refer to the file \texttt{covid.hh} of the available code.

\subsection{Part 3: ODE-Based Compartment Model} \label{sec:ODEModelDerivation}

As we have seen before, the federal states will be equipped with their own separate systems of equations. Hence, we start here by taking the system of PDEs obtained in the previous section, which was an approximation of a motion-based ABM, and reduce it further into a system of ODEs. 

We can imagine that, at this stage, Germany is divided into 16 domains, where each domain is represented by either an ABM or a PDE model. To reduce the PDE system further into a system of ODEs, some of these PDE federal states will soon become ODE federal states. This is why our notation changes here from $\Omega_{\text{PDE}}$ to $\Omega_{\text{ODE}}$. This reduction follows the procedure described in~\cite{Kehrer2025Hybrid}.

First, we take the derived weak formulation of the PDE system~(\ref{eq:PDEmodel}) with the integration over the soon-to-be ODE state $\Omega_{\text{ODE}}$ of Germany. 
We demonstrate this again using the equation for susceptibles with the susceptible density $S(x,t)$ and testing with the constant test function $\varphi=c$:
\begin{align*} 
    \int_{\Omega_{\text{ODE}}} \frac{\partial S}{\partial t} \varphi \; dx 
    &\approx 
    \int_{\Omega_{\text{ODE}}} - \nabla \varphi ^T D \nabla S  - \nabla \varphi ^T (S \nabla V) - \beta \pi r^2 S \bigl( I+\sY \bigr) \varphi \; dx \\
    %
    &\underset{\varphi = c}{=}  - \beta \pi r^2 c \int_{\Omega_{\text{ODE}}} S \left( I+\sY \right) \; dx.
\end{align*}
Further, let us decompose $S = \bar{S} + \tilde{S}$ into a spatially constant part $\bar{S} = \frac{1}{|\Omega_{\text{ODE}}|} \int_{\Omega_{\text{ODE}}} S \; dx$ and a fluctuating part $\tilde S$ with mean zero:
\begin{align*}
    \int_{\Omega_{\text{ODE}}} \frac{\partial \bar{S}}{\partial t} \varphi + \frac{\partial \tilde{S}}{\partial t} \varphi \; dx 
    &\approx 
    - \beta \pi r^2 c \int_{\Omega_{\text{ODE}}} 
     \underbrace{\bar{S} \left( \tilde{I}+\tilde{\sY} \right)
     +\tilde{S} \left( \bar{I}+\bar{\sY} \right)}_{=0 \text{ if integrated}} 
     + \bar{S} \left( \bar{I}+\bar{\sY} \right)
     + \tilde{S} \left( \tilde{I}+\tilde{\sY} 
     \right) 
    \; dx \\
    &= - \beta \pi r^2 c \int_{\Omega_{\text{ODE}}}  \bar{S} \left( \bar{I}+\bar{\sY} \right)
     + \tilde{S} \left( \tilde{I}+\tilde{\sY} 
     \right) 
    \; dx. 
\end{align*}
Assuming the domain is homogeneously mixed, the fluctuating part is approximately zero, and we get
\begin{align*}
     &&\frac{\partial \bar{S}}{\partial t} c |\Omega_{\text{ODE}}|
    &\approx \biggl(
     - \beta \pi r^2 c \bar{S} \left( \bar{I}+\bar{\sY} \right)
      \biggr)|\Omega_{\text{ODE}}|
      \\
 \Leftrightarrow && \frac{\partial \bar{S}}{\partial t}  
    &\approx - \beta \pi r^2 \bar{S} \left( \bar{I}+\bar{\sY} \right).
\end{align*}
We define $S_{\text{ODE}}$ to be the total number of susceptible individuals in the corresponding ODE domain $\Omega_{\text{ODE}}$. Hence, we multiply the constant part of the susceptible density $\bar{S}$ by the area size $|\Omega_{\text{ODE}}|$ to obtain the ODE 
for susceptibles in a single ODE federal state: 
\begin{align*}
    \dot{S}_{\text{ODE}} \approx 
    - \frac{\beta \pi r^2}{|\Omega_{\text{ODE}}|} S_{\text{ODE}} \left( I_{\text{ODE}}+\sY_{\text{ODE}} \right).
\end{align*} 
Similarly, we derive equations for the other compartments $E$, $I$, $\sY$, $H$, $C$, $H_C$, and $R$. The equations are similar to the PDE system~(\ref{eq:PDEmodel}). For implementation details, we refer to the file \texttt{covid-integrate.hh} of the available code.

\subsection{Initial Values} \label{sec:InitialValues}

We computed the 7-day average of Covid-19 data from the RKI \cite{rki2025SARS} 
for all federal states in Germany and used it to define the initial conditions of all model types. Given these transformed initial infection and recovery numbers, we set the \textit{exposed} ($E$) and \textit{recovered} ($R$) compartments of each non-ABM federal state to match the corresponding transformed data. Note that these are not necessarily integer values. 
For the ABM, this infection data were also used, but additionally scaled by a factor of 20 and rounded down to integer values. Due to stochasticity in the ABM, infections can die out more easily in the early phase. Therefore, higher initial values were chosen compared to the official RKI data. 
The initial numbers for the ODE and PDE contributions do not need to be scaled, as the corresponding models are deterministic. 
Besides the susceptible, exposed and recovered compartments, all other compartments were initially set to zero. The initial values for the susceptible compartment are then defined as the difference between the total population number for each federal state and the numbers of exposed and recovered individuals. This holds for each model type -- ABM, PDE, and ODE. 

The total population number was computed using the mobile phone data. All individuals are initialized based on their first recorded positions from the corresponding data. For the non-ABM domains, we store the initial population numbers, and for the ABM, we explicitly store the agent IDs. The initial population sizes of each federal state and the total population size of the simulation in Germany using the partial population sample (25\%) are given in Table~\ref{tab:populationSize}. 

\begin{table}[H]
    \centering
    \begin{tabular}{c||c}
         federal state & population size  \\
         \hline \hline
         Schleswig-Holstein & $680,266$ \\
         Hamburg & $341,928$ \\
         Lower Saxony & $1,934,288$ \\
         Bremen & $125,112$ \\
         North Rhine-Westphalia & $4,645,915$ \\
         Hesse & $1,615,589$ \\
         Rhineland-Palatinate & $1,056,170$ \\
         Baden-Württemberg & $2,762,181$ \\
         Bavaria & $3,450,204$ \\
         Saarland & $274,533$ \\
         Berlin & $711,327$ \\
         Brandenburg & $657,869$ \\
         Mecklenburg-Vorpommern & $368,422$ \\
         Saxony & $916,703$ \\
         Saxony-Anhalt & $536,096$ \\
         Thuringia & $539,874$ \\
         \hline \hline
         Germany & $20,616,477$
    \end{tabular} 
    \caption{Initial population size per federal state in Germany and total amount for the 25\% sample.}
    \label{tab:populationSize}
\end{table}
In the PDE component, the initial distribution of individuals is informed by the underlying landscape. To distribute the compartments (susceptible, exposed and recovered), we transform the landscape by taking into account the effect that the landscape of the PDE system has on the population during the simulation. This process is described in detail in our previous paper~\cite{Kehrer2026HybridAbmPde}. 

\subsection{Coupling} \label{sec:Coupling}

Agents from the ABM are not bound to their modeling domain. As our ABM is based on events, we first iterate through the events independently of the location. At the end of one ABM step, we store the agent IDs that should be transferred to a non-ABM domain so that these individuals are simulated during the next time step in a different domain. If an agent travels from one ABM domain to another ABM domain, nothing changes. For further information on when we decide to remove an agent, or how restrictions influence the coupling, see Section~\ref{sec:Implementation}. 
The leaving agent will either be added to the ODE domain or converted into the corresponding density amount and added to the PDE domain. How the distribution of each individual in a PDE domain is implemented explicitly is explained in detail in Section~\ref{sec:Implementation}.

While the transition from an ABM to a non-ABM domain is more or less straightforward, we wanted to keep the structure of how transitions are implemented in the other direction, so from non-ABM to ABM, as well as between non-ABM domains, as similar as possible. 
As stated previously, we imposed zero Neumann boundary conditions, ensuring that no individuals can enter or leave the PDE domain through its boundary. 
Furthermore, we preprocess the event data to explicitly track the agent IDs of individuals leaving a domain, the domain they enter, and the time step at which this transition occurs during the simulation. This enables a straightforward and realistic representation of transmission across the federal states in Germany.

\section{Implementation} \label{sec:Implementation}
The total population size of the 25\% sample in Germany is $20{,}616{,}477$ and stays constant throughout the entire simulation period at each time step when the systems are solved. 
We included an error message in case the entire population in Germany deviates from our initial total population number. All of our experiments ran without triggering this error. 
 
We computed the 7-day average of Covid-19 data from the RKI~\cite{rki2025SARS} 
for all federal states in Germany and used it as target data for evaluation, and for defining the initial conditions of all model types (see Section~\ref{sec:InitialValues}). 

As previously mentioned, we gathered information on each individual's initial location by iterating through the computed trajectories and assigned them to their corresponding federal state. 
Because infection cases in the RKI data depend on the residency of each individual, it makes sense to start the simulation at midnight, because most agents will be at home. For this reason, it would make sense to save the disease states only at midnight. However, we additionally store the disease-state data at each simulation time step to check the implementation. This allows us to verify that individuals are exchanged correctly between model domains and that the total population remains constant. Moreover, in our case, the mixing is not sufficiently high to produce strong temporal fluctuations. Therefore, we save the data for each time step. 

In addition to the original mobility extracted from the mobile phone dataset, the ABM incorporates mechanisms to simulate behavioral changes and policy interventions over time. One such mechanism is the use of activity change rates for out-of-home activities. These rates are defined on a daily basis and allow us to modify the original mobility data to reflect changes in social behavior, such as reduced movement during lockdowns (for details, see~\cite{Kehrer2026HybridAbmPde}). 
In an ABM domain, this is implemented by removing events. In a non-ABM domain, the activity reductions are implemented as a reduction in the infection rate~\cite{Kehrer2026HybridAbmPde}. 

Policy interventions such as school closures are implemented in a similar way to the activity reductions: if schools are marked as closed on a given day, all school-related events are excluded from the simulation for that period. School closures were not part of the simulation period considered in this paper. Nevertheless, the implementation of such interventions is included in the present framework and was previously used in our ABM-PDE model for the Berlin-Brandenburg region~\cite{Kehrer2026HybridAbmPde}.
Since school closure dates differed across the federal states and no uniform nationwide dates existed, we generally followed the information provided in~\cite{RND_datum_schulschliessungen} and chose dates between March 16--18, 2020 for all federal states in Germany.

The trajectories are identical for all weekdays (Monday to Friday) but differ between weekdays, Saturday, and Sunday. This is why we focus on a simulation period of one week when we compute the landscape, excluding activity restrictions, since these are random. 
Instead, activity restrictions are implemented differently in the PDE system. In principle, it would be possible to generate a different landscape for each fixed seed including activity reductions. If school closures were to be simulated, one could instead focus on an extended simulation period, since school closures can be implemented on fixed dates. 

The distribution of each individual in a PDE domain is not performed at a single spatial location, as this is unstable, but according to the current distribution, provided that this compartment is nonempty. If the compartment is empty, individuals are distributed according to the landscape, as done for the initial values.

Note that the federal state of Bremen consists of the two cities Bremen and Bremerhaven. For simplicity, we restrict our analysis to the city of Bremen only. Therefore, if a location falls within Bremerhaven, we instead assign it to the neighboring federal state of Lower Saxony. The initial infection numbers were distributed within Bremen alone. For a more accurate simulation, one would have to differentiate between both cities and initialize them separately. This can be done in our model by implementing them as separate domains.

To account for infectious and susceptible individuals who meet during the day, containers are created. Each container represents a facility of a certain category and exists for the current time step. It contains all relevant information needed to compute the exact duration of interaction between these individuals and to determine whether a susceptible individual becomes infected.
We update the health states during one ABM step. 
First, we go through every event and check whether an agent leaves a facility. Then, their health state may be updated. 
We save the new health states in a separate map, which is applied in the next time step, while accessing the old health states during the current update to avoid chain reactions.
Hence, different time step sizes influence the infection numbers and the overall infection dynamics.

We created a baseline scenario in which multiple federal states were represented by each model type. The data for the baseline scenario was created beforehand. For the experiments, we needed to create additional input data as we deviated from the baseline scenario. In general, the required data differs between model types. In \texttt{model\_setup.cpp}, one can move the federal states to the corresponding model type and adjust their corresponding indices. For ODEs, one needs the area size~\cite{citypopulation_germany} in $m^2$. 
For PDEs, one needs landscape data, its gradient, as well as grid information. 
For ABMs, one needs event data, facility data, the maximum number of agents per facility, and files containing initial agent IDs for each ABM domain. 
For all model types, initial infection data, initial population data, activity reduction data, 
and jump matrices with agent IDs are required. 
For our experiments, we changed Berlin from the baseline scenario model type "PDE" to both "ODE" and "ABM". For the change to "ODE", we only needed information on the area size. For the change to "ABM", we needed to preprocess additional data. Because we differentiated between non-ABM and ABM domains when computing the time point at which the transition occurs, the jump matrices also changed. This data had to be generated separately. However, most of the preprocessed data originates from the trajectories, which are independent of the chosen model types for the federal states.





Because all domains modeled using ABMs are combined into a single global ABM, these transitions do not need to be considered explicitly. All remaining transitions must be taken into account. The transitions between model types that need to be implemented are summarized in Table~\ref{tab:transitions}.

\begin{table}[!ht]
    \centering
    \begin{tabular}{c|c}
      removing individuals from state of model type &  adding individuals to state of model type \\
      \hline
       ODE & PDE \\
       PDE & ODE \\
       ODE & ODE \\
       ABM & ODE \\
       ODE & ABM \\
       ABM & PDE \\
       PDE & ABM \\
       (ABM) & (ABM) \\
       PDE & PDE 
    \end{tabular}
    \caption{Relevant transitions. Direct ABM-to-ABM transitions are not required, as all ABM federal states are simulated within a single global ABM.}
    \label{tab:transitions}
\end{table}
The question of how to determine the number of traveling individuals, or the number of transitions from one federal state to another based on the mobile phone data, is not straightforward to answer. While agents travel from one facility to another, they may enter not only a neighboring federal state but also a more distant one, and reaching their destination may take several hours. Since commuting agents can neither become infected nor infect others during their journey, we keep them in the ABM for as long as possible when their destination is a non-ABM federal state. Consequently, even if they left an ABM state several hours earlier, they remain in their originating ABM state until they reach their destination. This also means that we ignore any intermediate federal states they may pass through while commuting.
We chose this approach because, in a hybrid model, directly counting an agent as having left state A and entered state B in the next time step may result in the agent being transferred prematurely to the ODE or PDE model, where the individual immediately becomes active and can either transmit or acquire an infection. Furthermore, treating the movement through intermediate states as inactive removes the need to explicitly model and store the corresponding transition process.

Conversely, when individuals transition from a non-ABM domain to an ABM domain, we convert them into agents of the corresponding ABM domain as early as possible. For transitions between two federal states that are both represented by non-ABM models, we assign individuals to the federal state that is closest to their current position along the journey. Consequently, the selected federal state depends on the current time step as well as the departure and arrival times. With this approach, long-distance transitions can occur even between PDE domains that do not share a common border (see Fig~\ref{fig:germany_model_types}). 
Recall that transitions between ABM domains do not need to be performed explicitly, since all ABM federal states are simulated within a single global ABM. However, we still distinguish between individual ABM states when processing transitions into or out of non-ABM domains. At the end of each ABM step, we determine the federal state associated with each agent and store its current health state. 

We model transfers between non-ABM domains by distributing traveling individuals across health states according to the current compartment fractions in the source domain. Using fractional state representations for transfers reduces stochastic variance, particularly when the number of traveling individuals per time step is small. This approach is consistent with the continuous nature of ODE- and PDE-based domains, as these models describe populations in terms of proportions or densities rather than discrete individuals.
The amount of density removed from a PDE domain depends on the current density distribution over the grid. To determine the reduction in each compartment, we compute the compartment fractions and the total number of residents in the corresponding federal state. These quantities are then used to adjust the densities. 
Consequently, regions with higher densities, i.e., hotspots, are reduced more strongly than regions with lower densities. Similarly, when individuals enter a PDE domain, they are primarily added to regions where the density is already high.
If a compartment is empty but individuals still need to be added, they are distributed according to the landscape, as for the initial conditions. In contrast, a uniform reduction across the entire domain could lead to negative density values.

Transfers are not merged into net flows. If 300 individuals travel from Berlin to Brandenburg and 400 individuals travel from Brandenburg to Berlin, both transfers are performed explicitly rather than replacing them by a net transfer of 100 individuals from Brandenburg to Berlin. This approach is more realistic and preserves mixing between regions. Furthermore, it is required in the ABM, where the agent IDs of traveling agents are tracked explicitly. 

Note that activity reductions are not incorporated into the preprocessed jump matrix, since these reductions are applied randomly during runtime. If an individual in the ABM would normally transition to a non-ABM domain, but this transition is prevented by an activity reduction, the individual remains in the ABM and is therefore not removed from it. Consequently, instead of relying on the precomputed values stored in the jump matrix, we add only those agents that actually perform a transition to the non-ABM domain. As a result, the corresponding entries in the jump matrix are no longer accurate, but they are not used in this case.
This modeling choice is computationally more expensive, since the location of each agent at the end of the time step must be determined explicitly rather than obtained from precomputed transition information. However, without this procedure, the transition numbers would be inaccurate and the effect of the activity reductions would only be partially represented.
Before adding agents to the ABM and removing them from a non-ABM domain, we additionally verify whether these agent IDs already exist in the ABM as a consequence of activity reductions. By accounting for these rare cases, we avoid introducing duplicate individuals into different federal states. 
To account for travelers who visit multiple federal states, this check is performed for the agent IDs of all ABM domains. For transitions between non-ABM domains, activity reductions do not affect the number of traveling individuals. 

Similarly to the previous papers, the boundary grid nodes of each modeling state were obtained from OpenDataLab~\cite{RandwerteDaten}. Using these data, we visualized our results across Germany and created triangulated domains for the PDE states.   
We used Triangle~\cite{triangle} to create the triangulations of the corresponding PDE domains. The PDE systems of our hybrid model were solved using the finite element method implemented in the Kaskade7 software~\cite{GoetschelSchielaWeiser2020}, with the Dune interface. Correspondingly, we implemented our ABM and ODE systems in C++ as well. 
For pre- and postprocessing of the data, as well as for the visualization of our simulation results, we used Python~3.9 and, in particular, the \texttt{matplotlib} package for plotting. For the projection from geographical coordinates to UTM coordinates in Python, we used the function \textit{Proj} from the package \textit{pyproj} with the coordinate reference system \textit{EPSG:25832}. 

\begin{figure}[h!]
    \centering
    \includegraphics[width=0.55\linewidth]{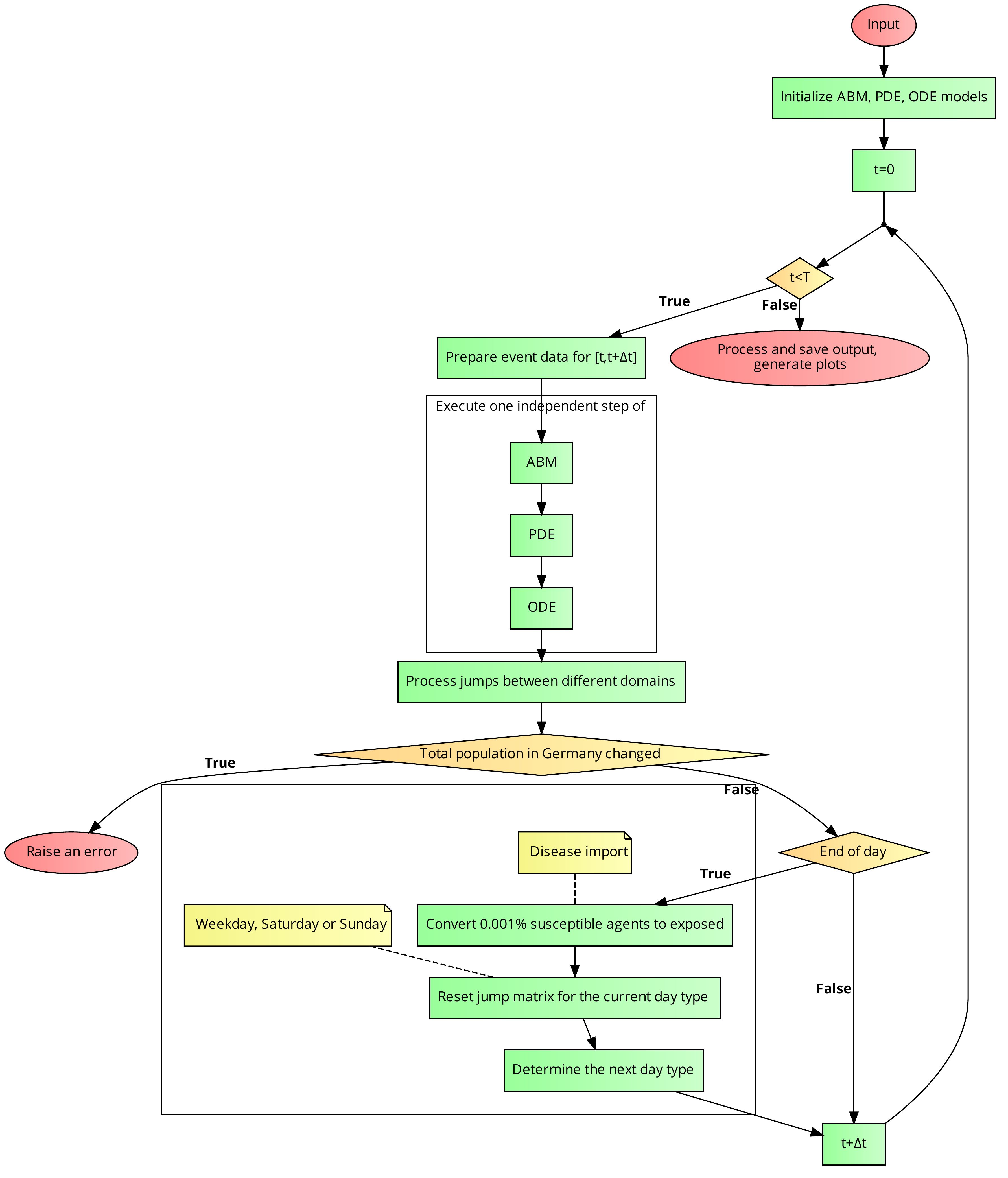} 
    \caption{Flowchart}
    \label{fig:flowchart}
\end{figure}

We generally performed 10 simulation runs using 10 fixed but different seeds, which were executed in parallel. Furthermore, the ABM, PDE, and ODE steps could also be parallelized, since they are executed independently. However, this additional level of parallelization was not implemented in the present work. 
An overview of the simulation workflow is provided in Fig~\ref{fig:flowchart}.

\noindent
\section{Choice of Transition Rates} 
\label{sec:ParameterOptimization}


In the hybrid framework, the underlying infection mechanisms differ between the ABM and the deterministic models. 
The infection rate for a single susceptible individual in the ABM is determined by three components: 
(i) a baseline calibration parameter, adopted from EpiSim and scaled by a constant factor to account for the simplified implementation used here, 
(ii) the contact intensity, and 
(iii) the cumulative interaction duration with infectious individuals during the time step. A more detailed description of the infection model and parameter formulation can be found in~\cite{Muller2021ABM,Kehrer2026HybridAbmPde}. 

For the deterministic model components, most transition rates can be transferred directly from the ABM, since they describe disease progression rather than contact-dependent transmission. We therefore adopt all transition rates from the ABM for the deterministic models, except for the infection rate, because the infection process depends on several heterogeneous contact-related parameters that may not be reliably represented by a single average value. Note that we did not attempt to use an averaged infection rate, although this could be explored in future research. 
We adopted the optimized parameter values from our previous paper~\cite{Kehrer2026HybridAbmPde} as initial parameter values. The optimized infection rate determined for the PDE framework was applied uniformly to all PDE and ODE states, while the optimized infection rate obtained for the ABM framework was applied uniformly to all ABM states. In the current paper, however, the number of federal states considered is substantially larger than in the previous work. We assumed that additional optimization would be necessary, not only because of the higher probability that a traveler is infected (due to fractional transitions and because the population outside the modeled region is now active), but also because every federal state may have its own infection dynamics and we never optimized an infection rate for an ODE state. 

The previous paper incorporated commuting individuals across all of Germany, which implies that the initial parameter choices should not deviate substantially from optimized values, at least for Berlin and Brandenburg. For consistency, we therefore adopted the same correction steps as in the previous paper: 0.05 for the ABM and 50.0 for the PDE (and ODE) states. In the present setting, each state is assigned its own infection rate. For each parameter configuration, we conducted 10 simulation runs using fixed seeds. 

After evaluating the initial runs, the ODE-state results showed deviations from the reference results, indicating the need for higher infection rates. 
During the 
evaluation process, the error was measured by averaging the results for each state and subtracting the corresponding target data. Summing these differences over all days revealed whether the results over- or underestimated the target data. 
We therefore manually selected higher infection rates for all ODE states, resulting in errors that were acceptable for the proof-of-concept setting. We ended up with infection rates of $0.4 \cdot 1.7e-5$ for all ABM states, $4.5e+2$ for all PDE states and 
$5.45e+3$ for all ODE states in the baseline scenario. 
Although further parameter optimization could reduce the error, the current parameter choices are sufficient for demonstrating the functionality of the framework.
%

\section{Experiments}

In this section, we evaluate the proposed hybrid framework from several perspectives. First, we discuss the assignment of model types to the federal states of Germany (Section~\ref{subsec:ModelTypes}). We then investigate the trade-off between computational cost and accuracy by representing Berlin as an ABM, PDE, or ODE domain (Section~\ref{subsec:single_state_different_models}). Next, we analyze the effect of complete border closures on the epidemic dynamics (Section~\ref{subsec:closing_borders}). Finally, we implement 
and compare Zero-COVID and No-COVID intervention strategies and assess their effectiveness in reducing the number of symptomatic individuals (Section~\ref{subsec:zero_no_covid}). 

\subsection{Choice of Model Types for Each State} 
\label{subsec:ModelTypes}

Before selecting a model type for each federal state, we must consider which properties make a specific model computationally expensive or otherwise unsuitable. The implementation allows each federal state to be assigned any of the available model types: ABM, PDE, or ODE. 
Since this work focuses on a proof-of-concept model, the choice of model type for each state is made primarily based on runtime considerations. Before presenting these choices, we briefly discuss general criteria that may influence the suitability of different model types.

Among the three model types considered here, the ABM provides the highest level of detail, as it explicitly represents individuals and incorporates high-resolution mobility data. Furthermore, transitions between federal states are based on mobility data with explicit agent IDs. Their timing depends on the model types involved, with individuals remaining in ABM domains for as long as possible to avoid an early contribution to the infection dynamics of the reduced models. Replacing the ABM representation of selected federal states with our reduced models introduces several simplifications: (1) deterministic health state progression, (2) the lack of information about previous individual health states in non-ABM domains, (3) the representation of activity reductions through a reduced infection rate, (4) the loss of spatial resolution, and (5) the aggregation of individuals into population-level quantities.

These aspects can be assessed individually. 
The first point is primarily relevant for small population sizes. For sufficiently large populations, stochastic fluctuations at the individual level have little impact on epidemic dynamics. 
Close agreement between a stochastic ABM and a corresponding deterministic ODE model has been reported for populations of approximately 20,000 individuals~\cite{AlAbri2025Analysing}. In contrast, small populations may exhibit discrepancies due to stochastic extinction and the discrete nature of ABMs. To assess the relevance of this effect for our model, additional simulations would be necessary. It is noteworthy that even the smallest federal state considered in our study, Bremen, has a population of 125,112 individuals (see Table~\ref{tab:populationSize}), well above this threshold, suggesting that stochastic discrepancies between model types are unlikely to be a major concern in any of the federal states considered here. However, stochastic models can provide not only a single expected trajectory, but a spectrum of possible epidemic outcomes across repeated simulation runs. This information is lost when using deterministic models. 
The second point is left for future work, as it can be addressed by extending the current framework. Since we consider only an early phase of the epidemic, most, if not all, transitioning individuals are susceptible, so no meaningful health-state history is lost when agents cross into a non-ABM domain. Therefore, the effect is currently expected to be negligible.
The third point concerns activity reductions. In the ABM, these are implemented by removing activities that take place outside the home. When an event is removed, the individual instead commutes directly to the next destination at the originally scheduled arrival time. As a result, the commuting period is extended, which prevents potential infection transmission during the time of the omitted event.
This effect can be translated into a reduced infection probability. However, the exact relationship is unclear, as omitted events may last from only a few minutes to several hours. Further study is required to establish a relation between event removal in the ABM and a corresponding reduction in the infection rate within the ABM, as well as its translation to an appropriate reduction in the infection rate of non-ABM models.
The fourth point is primarily relevant when spatial heterogeneity within a federal state significantly influences the epidemic dynamics. This is especially important when interventions such as school closures are considered. Since the simulation period is relatively short, no school closures are considered. In our previous work~\cite{Kehrer2026HybridAbmPde}, we addressed this issue by fitting the infection rate separately for simulation periods with and without school closures. 
Additionally, we consider artificial interventions, such as border closures, which are implemented by restricting travel between federal states while leaving mobility within each federal state unchanged. When commuting is not considered, less detailed model types may become preferable, since the implementation of mobility between model domains, in particular the timing of transitions, no longer plays a role. We also implement interventions based on case tracking, such as No-COVID and Zero-COVID, which are more difficult to represent in deterministic models because individual infections cannot be tracked explicitly and compartments generally do not become empty (except initially or after sufficiently long simulation periods). In the ABM domains, agents remain at home while restrictions are active. In the non-ABM domains, we again employ activity reductions by reducing the infection rate. 
Furthermore, federal states cover large geographical areas and therefore may exhibit substantial variation in population density and mobility patterns. As a result, epidemic dynamics may differ considerably between regions within the same federal state. Such spatial heterogeneity can be represented by PDE or metapopulation models, whereas an ODE model assumes a well-mixed population at the scale of the entire federal state.
Finally, the fifth point concerns the loss of individual identities in the reduced models. This aspect is closely related to the first point, since individual-level effects may become less relevant in sufficiently large populations. However, explicit individual representation can still simplify the implementation of mechanisms that depend on individual properties, or local contexts. For example, ABMs can represent room sizes, air quality, contact durations, or repeated exposure patterns more directly, making infection transmission more traceable at the individual level. They also allow interventions to be applied only to specific parts of the population or under certain conditions, such as mask wearing.

\begin{figure}[h!]
    \centering
    \includegraphics[width=0.6\linewidth]{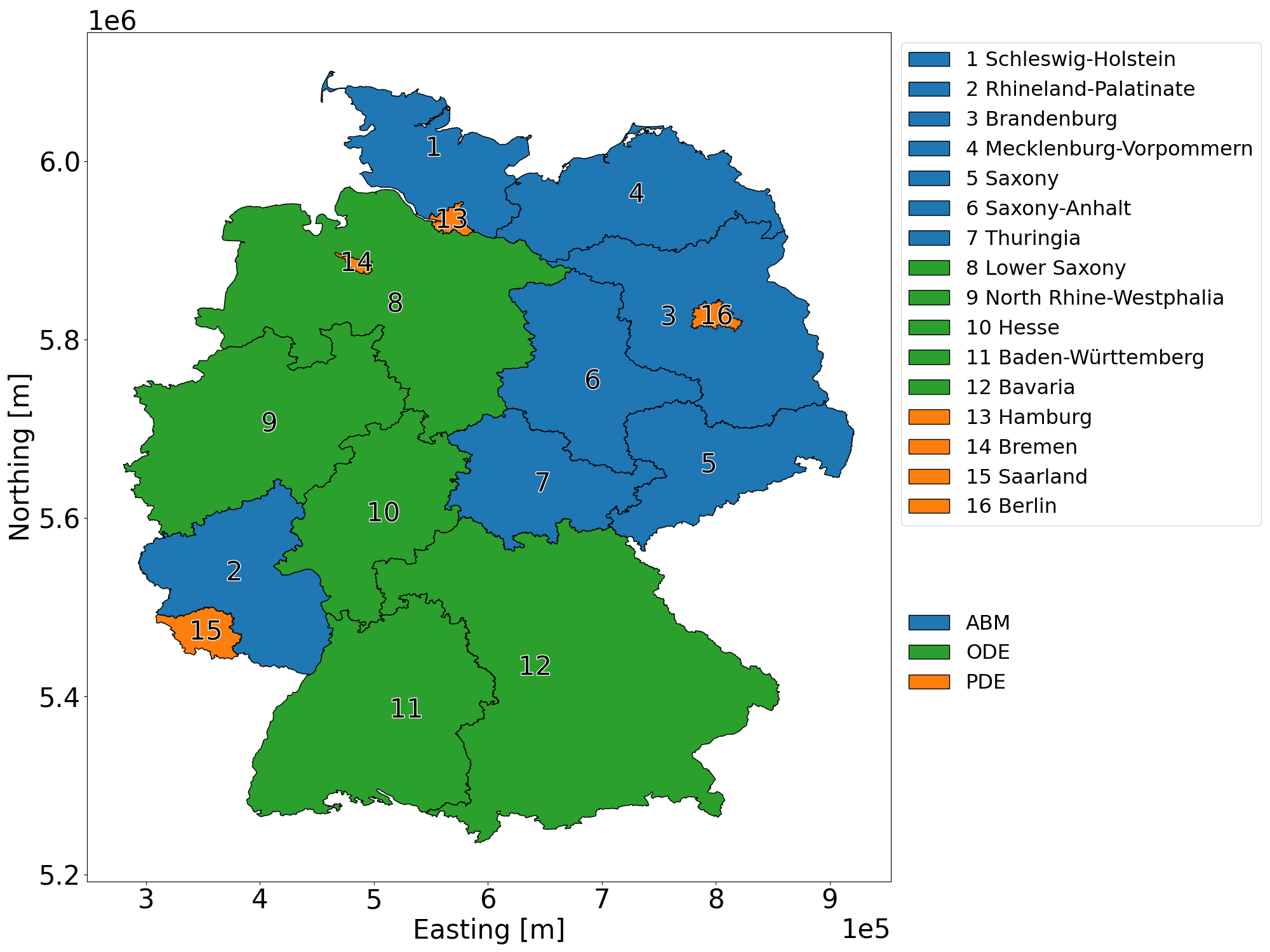} 
    \caption{Boundary data of all states in Germany. The chosen color indicates the corresponding model type.}
    \label{fig:germany_model_types}
\end{figure}

On the other hand, reduced model types generally offer lower computational costs and shorter runtimes. The ABM can become computationally expensive for several reasons. A large number of agents increases the overall computational load. Highly active agents are associated with many events, further increasing the runtime. In addition, having many susceptible individuals co-located with numerous infectious $I$ or symptomatic agents $\sY$ in the same container increases the number of interactions that must be processed. Frequent jumps of agents between regions also require additional bookkeeping and updates, which can further increase the computational cost.
In contrast, PDE simulations are primarily affected by the number of nodes in the grid, making a fine spatial resolution computationally expensive. Therefore, it can be advantageous to represent smaller regions using PDEs rather than ABMs. 
ODEs are generally the fastest model type to simulate, making them suitable for efficiently representing the remaining federal states. 
When ordering the federal states by population size, a pronounced gap becomes apparent, providing a natural threshold of 5 million individuals (or 1.25 million individuals for 25\% of the population) for distinguishing between ABM and ODE representations. Fig~\ref{fig:germany_model_types} shows the model type assigned to each federal state. 




\subsection{Single State - Different Models} \label{subsec:single_state_different_models}

\begin{figure}
    \centering
    \includegraphics[width=0.5\linewidth]{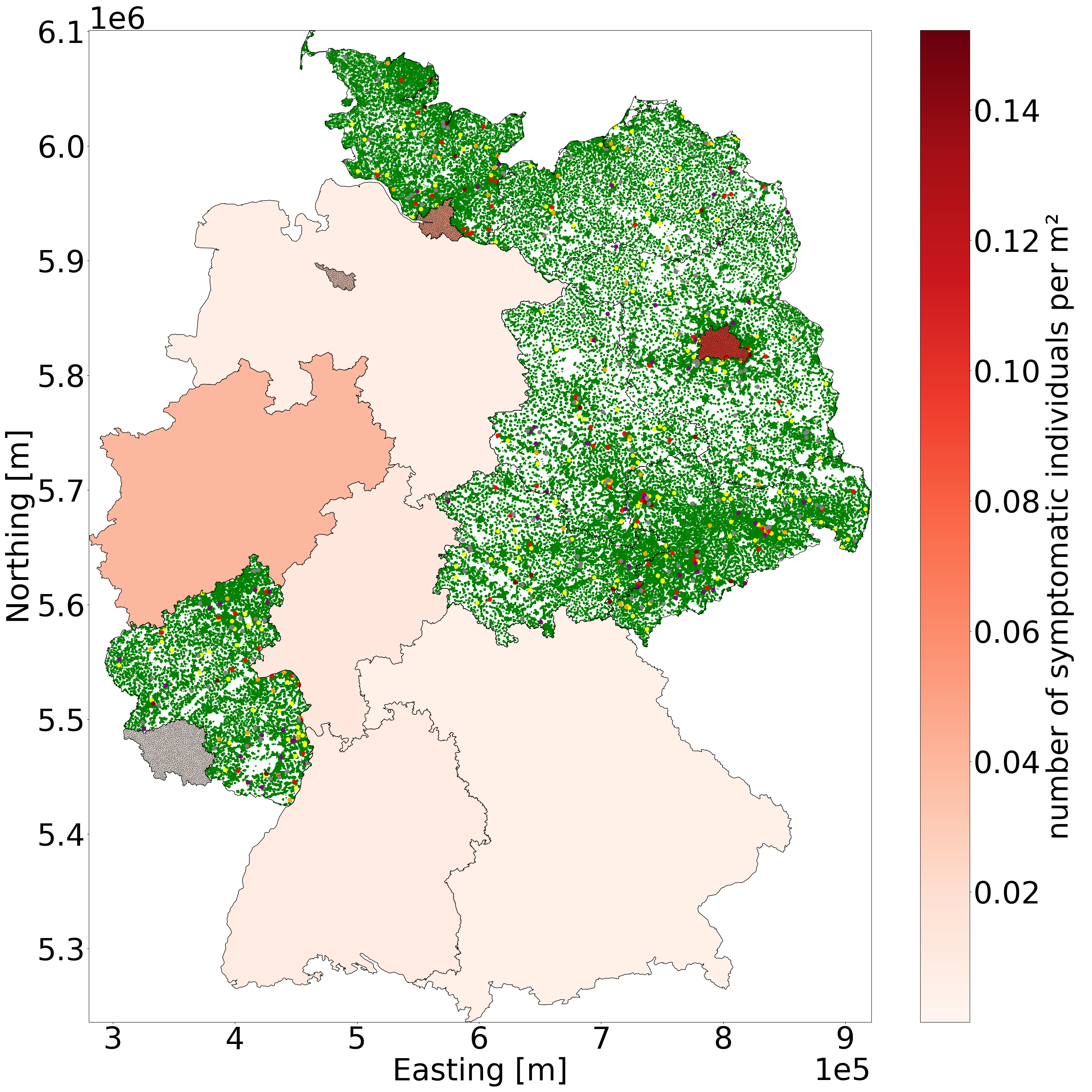}
    \caption{
    Visualization of all model types in Germany for a single simulation run. Shown are the symptomatic density in ODE and PDE federal states and the agents in ABM federal states on day 12 at 12:30 PM. Agent health states are indicated by different colors:
    \textcolor{softgreen}{susceptible}, 
    \textcolor{yellow}{exposed},
    \textcolor{orange}{infectious}, 
    \textcolor{red}{symptomatic}, 
    \textcolor{darkred}{requires hospitalization}, 
    \textcolor{purple}{critical}, 
    \textcolor{gray}{recovered}.}
    \label{fig:agents_and_density}
\end{figure}

\begin{figure}[h!] 
    \centering
    \begin{subfigure}[b]{0.33\textwidth}
       \includegraphics[width=\linewidth]{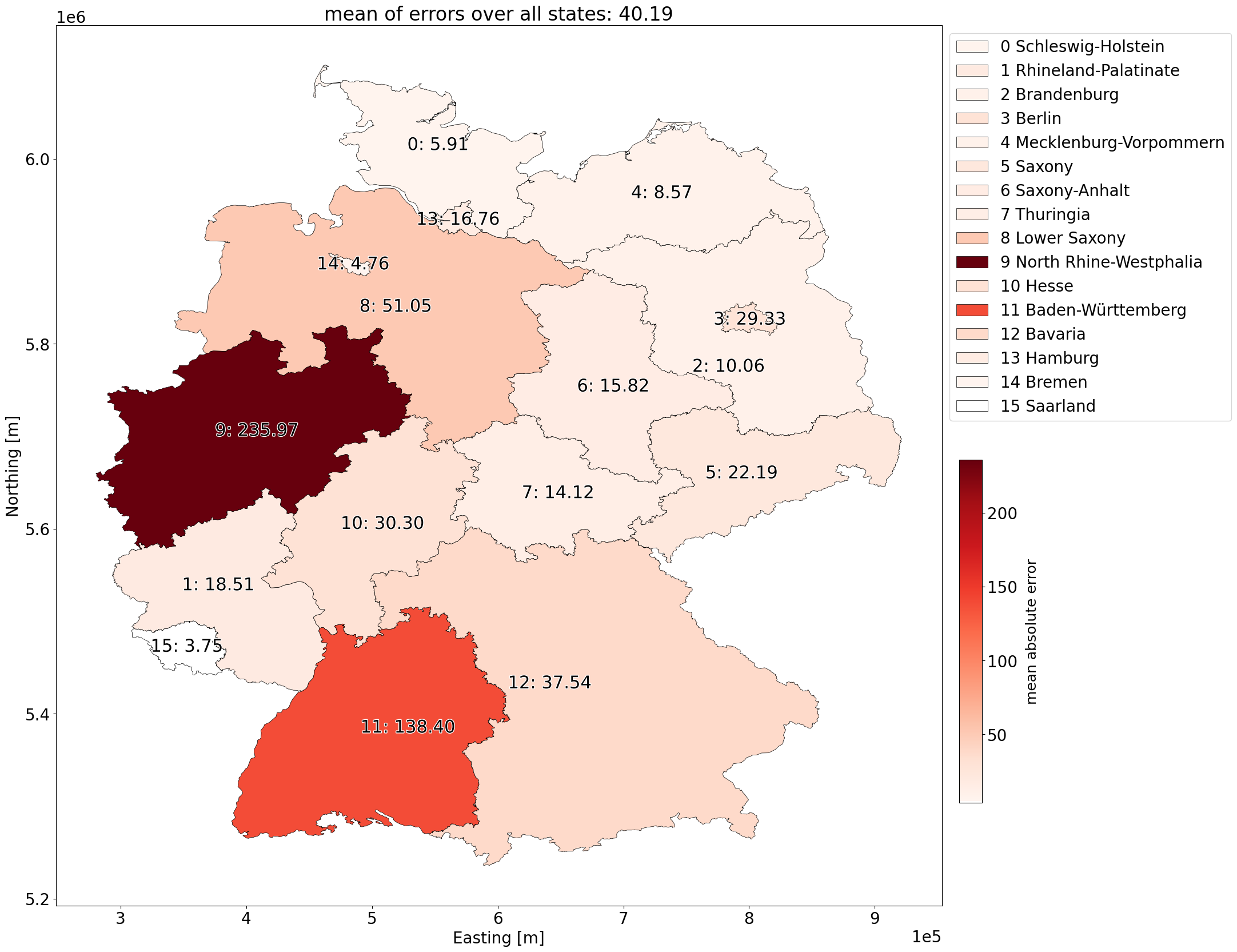}
        \caption{Berlin as ABM}
        \label{fig:error_Berlin_as_ABM}
    \end{subfigure}
    \begin{subfigure}[b]{0.33\textwidth}
       \includegraphics[width=\linewidth]{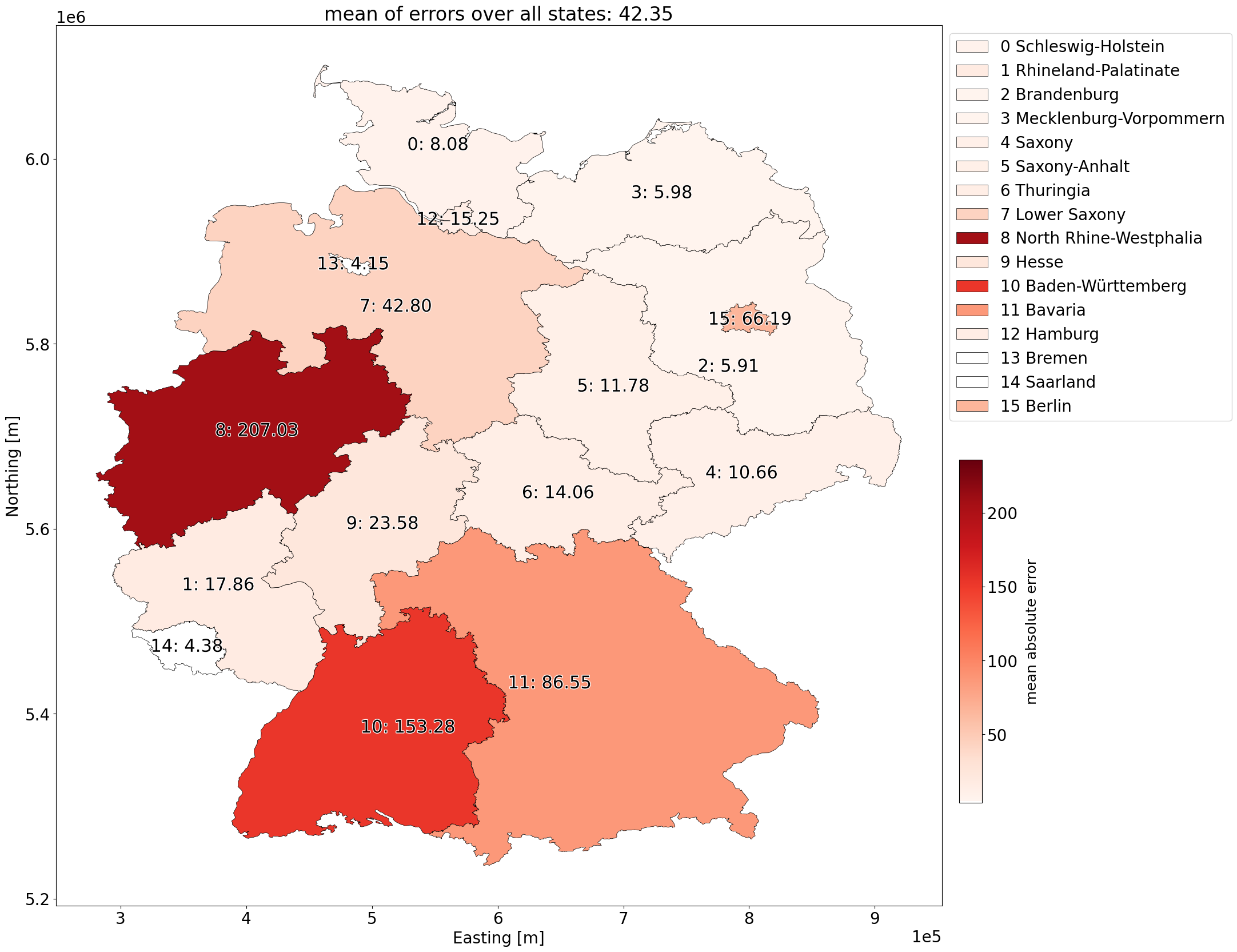}
        \caption{Berlin as PDE}
        \label{fig:error_Berlin_as_PDE}
    \end{subfigure}
    \begin{subfigure}[b]{0.33\textwidth}
       \includegraphics[width=\linewidth]{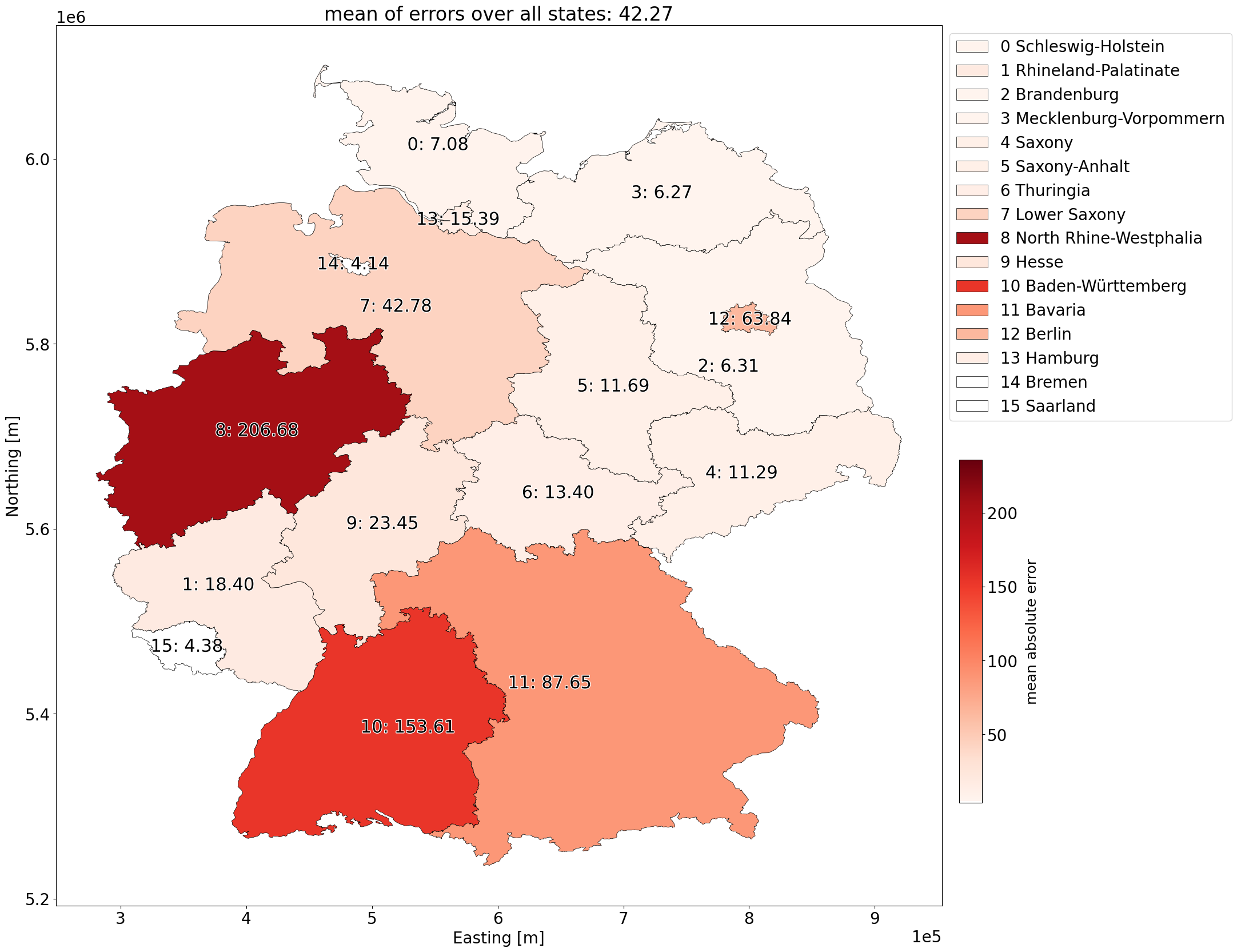}
        \caption{Berlin as ODE}
        \label{fig:error_Berlin_as_ODE}
    \end{subfigure}
    \caption{Average error across the federal states of Germany over 10 simulation runs: simulated number of symptomatic individuals compared with the daily target data.}
    \label{fig:error_Berlin_as_ABM_PDE_ODE}
\end{figure}

We model a single federal state using an ABM, PDE, and ODE representation and compare the resulting computational costs and accuracy. For these experiments, we selected Berlin. The baseline scenario with Berlin represented as a PDE domain is visualized in Fig~\ref{fig:agents_and_density} for a single simulation run at a fixed time step. The figure shows the symptomatic density in the non-ABM federal states as well as the locations and health states of the agents in the ABM federal states. The errors for each federal state and for Germany as a whole are shown in Fig~\ref{fig:error_Berlin_as_ABM_PDE_ODE} for all three model configurations. To facilitate comparison across the configurations, the same scale is used for all plots. 
As mentioned previously, further parameter optimization could improve the results. In particular, the infection rates were not re-optimized for the different model representations. Instead, Berlin retained the same correction term for the infection rate in the ABM and ODE configurations as in the PDE configuration. 
Since the target numbers are relatively small, the mean absolute error (MAE) is a more informative performance measure than the mean relative error (MRE).

Among the considered model configurations, the ABM representation of Berlin yields the lowest overall error for Germany, followed by the ODE and PDE representations. The same ranking is observed for Berlin itself.

Although the correction terms for the infection rate were identical across all model configurations, the predicted dynamics in all federal states vary depending on how Berlin is represented. 
However, the presented results are averages over only ten simulation runs, which limits the conclusions that can be drawn. 
Although there is no difference in the number of jumps between the PDE and ODE representations, the choice of an ABM representation does have an effect. Since individuals remain in the ABM domain for as long as possible before transitioning between ABM and non-ABM domains, simulations with a larger number of ABM domains that are highly connected to non-ABM domains can capture mobility patterns more realistically. If a non-ABM domain that has no connections to other non-ABM domains were replaced by an ABM domain, the representation of mobility within the transition process would therefore remain essentially unchanged. In practice, however, the German federal states are strongly interconnected, so the assigned model types remain relevant for the transition process. 
This effect is particularly relevant for federal states with high commuting volumes. Therefore, it may be beneficial to represent federal states with substantial commuter traffic using an ABM.
%
The runtimes show that the hybrid ABM-PDE-ODE model with Berlin represented as an ABM is indeed the slowest, whereas the model with Berlin represented as an ODE is the fastest (see Table~\ref{table:times}). The runtime reduction is much larger when replacing the ABM representation of Berlin by a PDE representation than when replacing the PDE representation by an ODE representation. However, this observation is specific to the choice of model types and federal states considered in this work and cannot be expected to hold in general. The overall runtime is also influenced by the additional costs associated with coupling different model types. Furthermore, while the computational cost of ODE models is independent of the area size and population size of a federal state, this is generally not the case for ABM and PDE models.

\begin{table}[!ht]
    \centering
    \begin{tabular}{c||c|c|c}
    Berlin as              & ABM     & PDE & ODE \\
    \hline
    mean runtime & $124.85$ & $112.37$ & $111.54$ \\
    absolute difference to previous configuration & -- & $12.48$ & $0.83$
    \end{tabular}
    \caption{Mean runtimes in minutes (min.) over 10 simulation runs for nationwide simulations of Germany using different model configurations.}
    \label{table:times}
\end{table}
The framework also allows the investigation of intervention strategies that restrict movement between regions. While such large-scale restrictions may be difficult to implement in practice for economic and societal reasons, we nevertheless consider several restriction scenarios in the following two subsections to further evaluate the model behavior.
%
\subsection{Closing Borders} \label{subsec:closing_borders}

To assess the influence of an individual federal state on the infection dynamics, we consider a scenario in which all travel into and out of a selected federal state is prohibited for the entire simulation period. 

In the ABM, border closures are implemented by suppressing transfers between federal states whenever they would occur according to the mobility data. This differs from the activity reduction measures discussed previously, as mobility events are not removed. Instead, agents continue to participate in their scheduled activities, but they remain associated with their original federal state and are not transferred into the model representation of the restricted state. This is consistent with the standard treatment of short-term travel in our hybrid model, where agents may visit locations in other federal states during a time step while remaining formally represented in the ABM at the beginning and end of the time step. Consequently, agents from other federal states may still travel to locations within the restricted state and interact with one another there, but they do not interact with individuals belonging to the restricted state itself. In this sense, the restricted state and the visiting agents are represented by separate, non-interacting populations occupying the same geographical area. From a real-world perspective, this can be interpreted as relocating the corresponding activities to alternative locations within the agents' home state, allowing them to maintain their usual contacts while preventing interactions across the closed border.

When restrictions are imposed, agents continue to participate in their scheduled activities and may still contribute to local transmission dynamics within their own population. For federal states represented by deterministic models, such as the PDE or ODE models considered here, the epidemic dynamics within the restricted federal state are independent of the remaining federal states once all incoming and outgoing travel is prohibited. Consequently, the solution inside the restricted state is identical across simulation runs and only needs to be computed once if its epidemic dynamics are of interest. Otherwise, the restricted state can be excluded from the simulation entirely, reducing computational costs.

\begin{figure}[h!] 
    \centering
    \begin{subfigure}[b]{0.33\textwidth}
       \includegraphics[width=\linewidth]{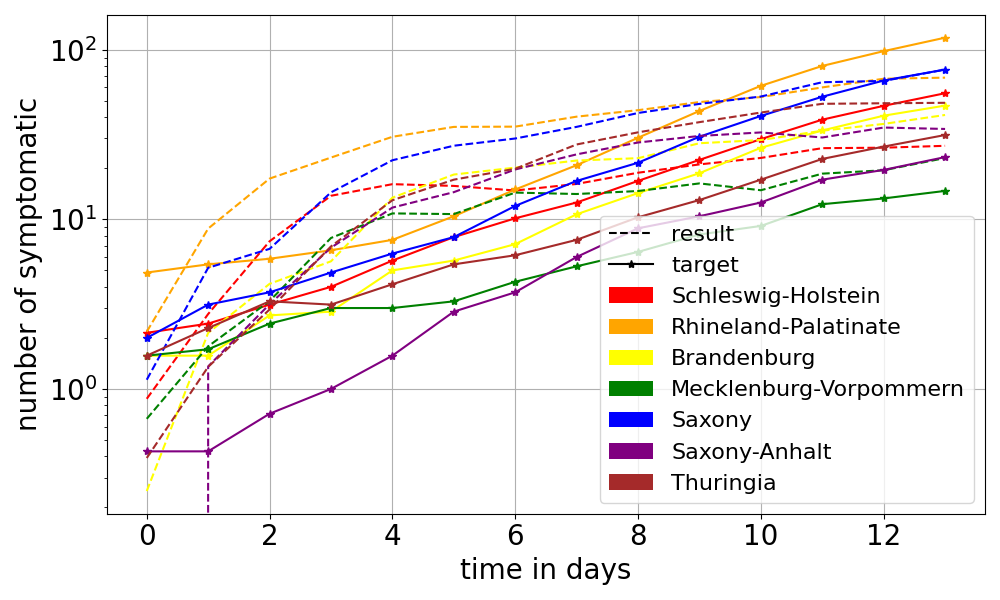}
        \caption{ABM federal states}
        \label{fig:result_ABM_Berlin_as_PDE}
    \end{subfigure}
    \begin{subfigure}[b]{0.33\textwidth}
       \includegraphics[width=\linewidth]{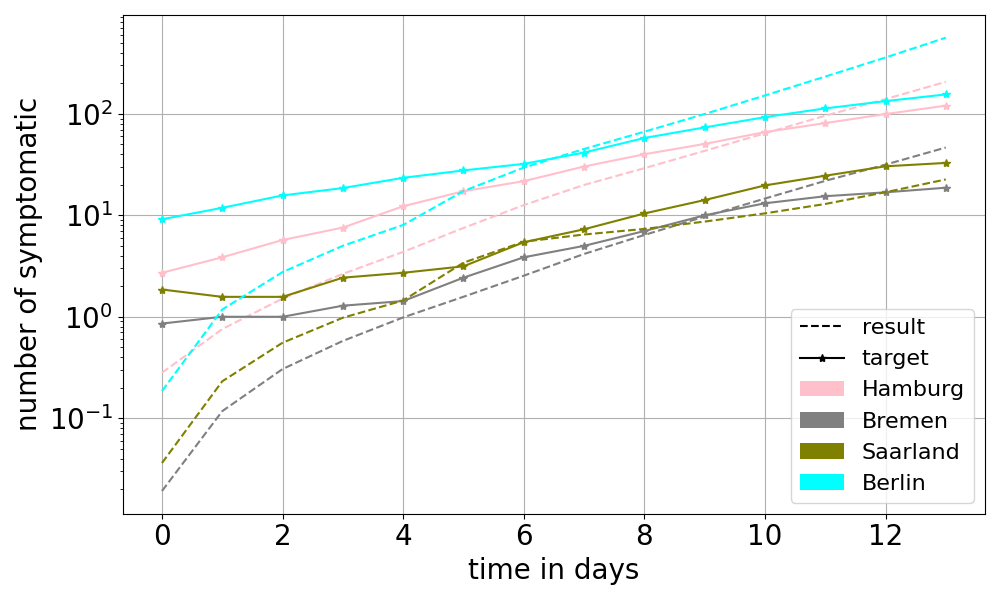}
        \caption{PDE federal states}
        \label{fig:result_PDE_Berlin_as_PDE}
    \end{subfigure}
    \begin{subfigure}[b]{0.33\textwidth}
       \includegraphics[width=\linewidth]{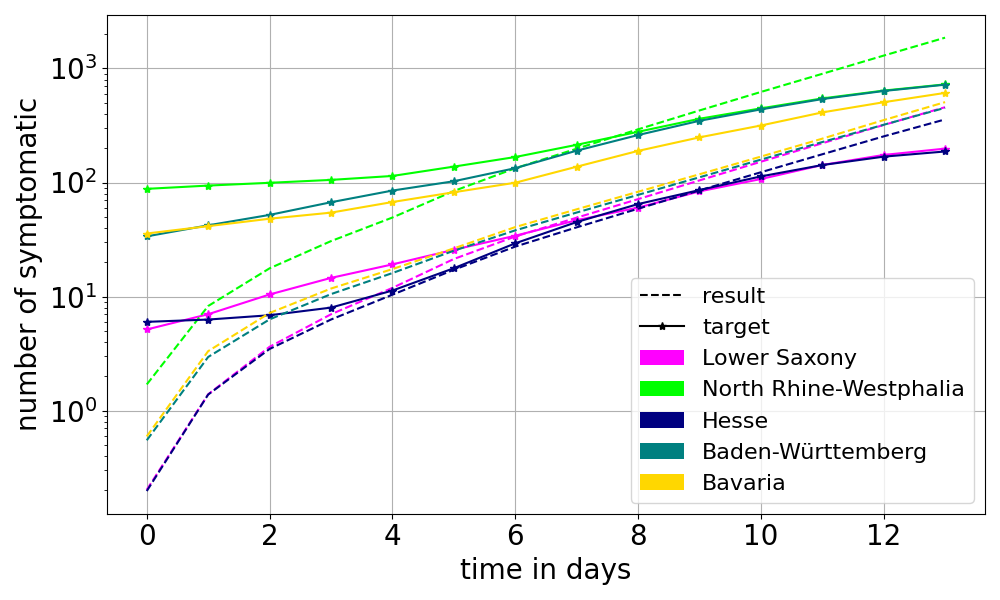}
        \caption{ODE federal states}
        \label{fig:result_ODE_Berlin_as_PDE}
    \end{subfigure}
    \caption{Results for the baseline scenario across the federal states of Germany, averaged over 10 simulation runs: simulated number of symptomatic individuals compared with the daily target data.}
    \label{fig:result_Berlin_as_PDE}
\end{figure}

\begin{figure}[h!] 
    \centering
    \begin{subfigure}[b]{0.33\textwidth}
       \includegraphics[width=\linewidth]{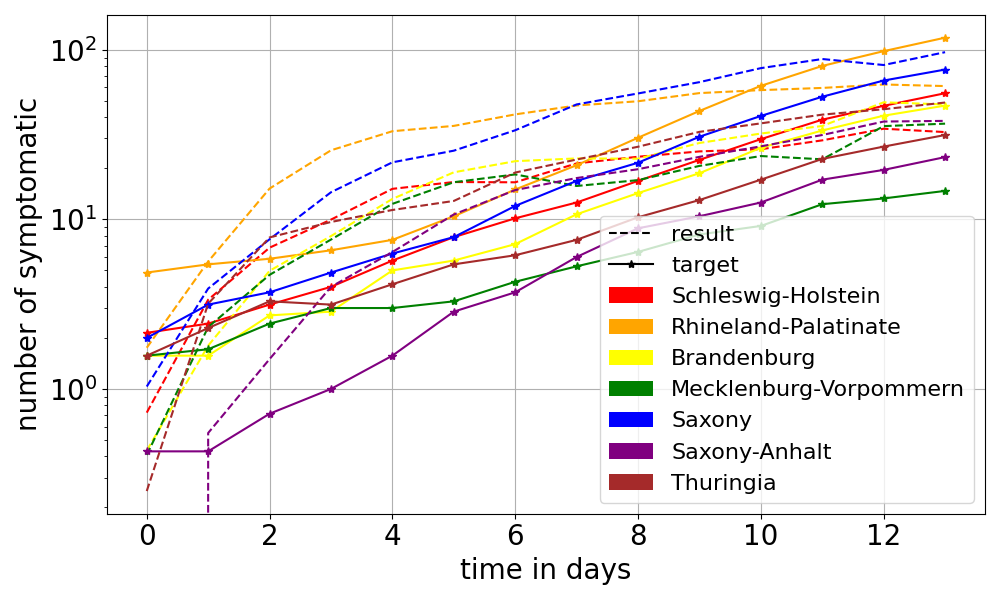}
        \caption{ABM federal states}
        \label{fig:result_ABM_without_Berlin}
    \end{subfigure}
    \begin{subfigure}[b]{0.33\textwidth}
       \includegraphics[width=\linewidth]{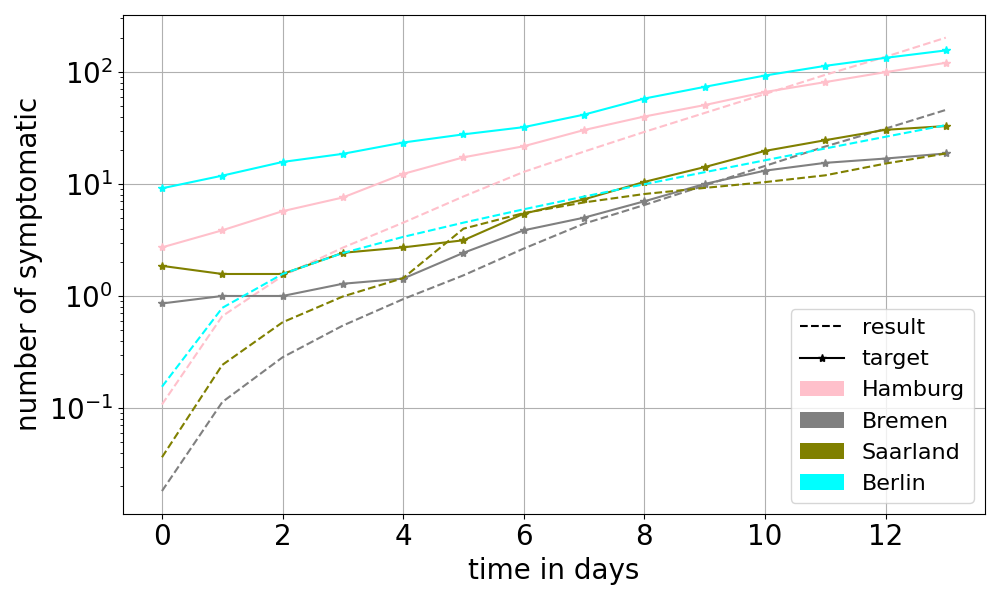}
        \caption{PDE federal states}
        \label{fig:result_PDE_without_Berlin}
    \end{subfigure}
    \begin{subfigure}[b]{0.33\textwidth}
       \includegraphics[width=\linewidth]{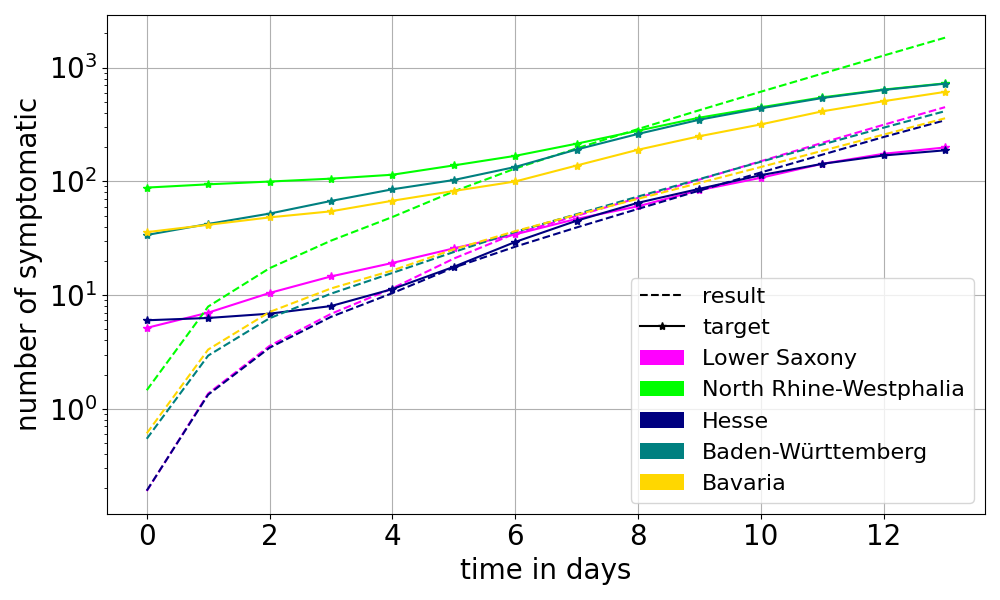}
        \caption{ODE federal states}
        \label{fig:result_ODE_without_Berlin}
    \end{subfigure}
    \caption{Results for the baseline scenario with total travel restrictions to and from Berlin across the federal states of Germany, averaged over 10 simulation runs: simulated number of symptomatic individuals compared with the daily target data.}
    \label{fig:result_without_Berlin}
\end{figure}

As an example, we apply a complete border closure to Berlin, which is represented as a PDE domain. The results, averaged over 10 simulation runs, are shown in Fig~\ref{fig:result_without_Berlin}. For comparison, the corresponding results of the baseline scenario are shown in Fig~\ref{fig:result_Berlin_as_PDE}. The mean numbers of symptomatic individuals for both scenarios, as well as for additional scenarios discussed later, are summarized in Table~\ref{tab:border_closure}. 

\begin{table}[!ht]
    \centering
    \begin{tabular}{c||c|c|c|c}
        border closure & none & Berlin & NW &  all \\
        \hline
        mean number of symptomatic individuals  &  $1175$ 
        & $1039$ 
        & $1447$ 
        & $1047$ 
        \\
        absolute difference to baseline scenario            & $0$ & $136$ & $-272$ & $128$ \\
        mean number of symptomatic individuals without NW   & $752$ & $623$ & $547$ & $147$ \\
        absolute difference to baseline scenario without NW & $0$ & $129$ & $205$  & $605$ 
    \end{tabular}
    \caption{Comparison of border-closure scenarios. Mean number of symptomatic individuals in Germany for the baseline scenario (none) and for border closures of North Rhine-Westphalia (NW), Berlin, and all federal states, averaged over 10 simulation runs and the 14-day simulation period. Values are rounded to the nearest integer.}
    \label{tab:border_closure}
\end{table}
\begin{figure}[h!] 
    \centering
    \begin{subfigure}[b]{0.33\textwidth}
       \includegraphics[width=\linewidth]{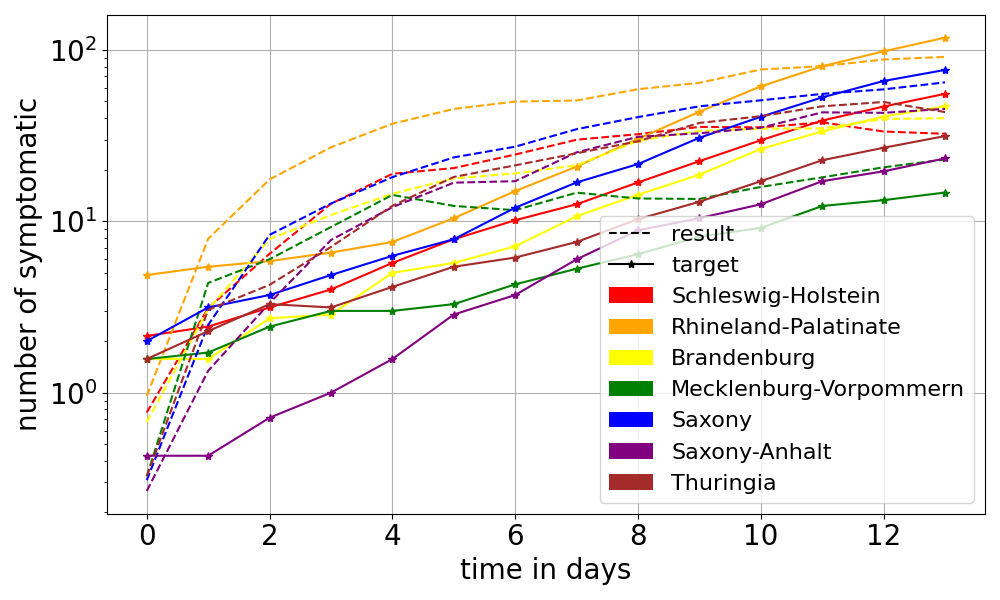}
        \caption{ABM federal states}
        \label{fig:result_ABM_without_NW}
    \end{subfigure}
    \begin{subfigure}[b]{0.33\textwidth}
       \includegraphics[width=\linewidth]{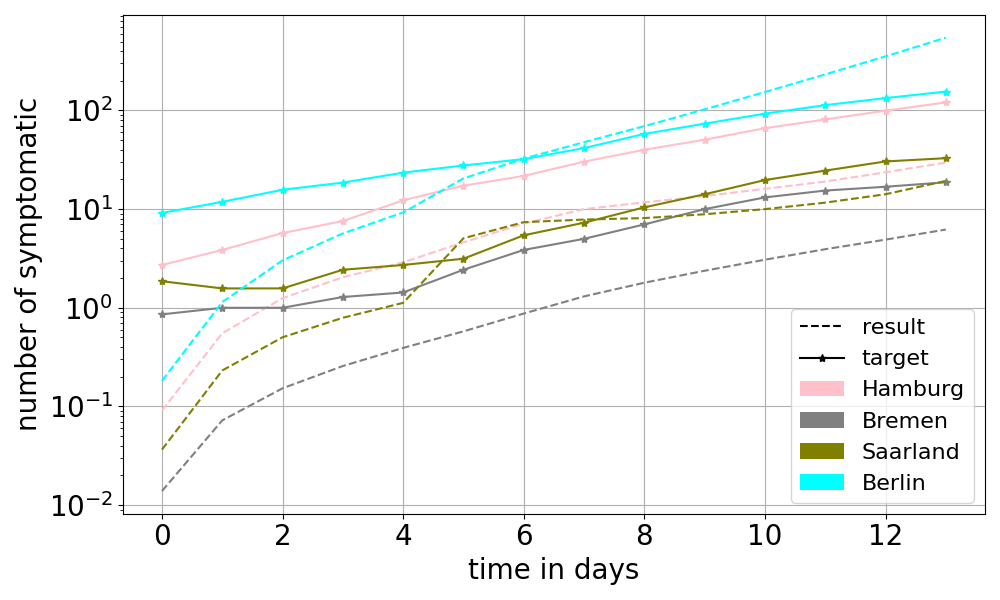}
        \caption{PDE federal states}
        \label{fig:result_PDE_without_NW}
    \end{subfigure}
    \begin{subfigure}[b]{0.33\textwidth}
       \includegraphics[width=\linewidth]{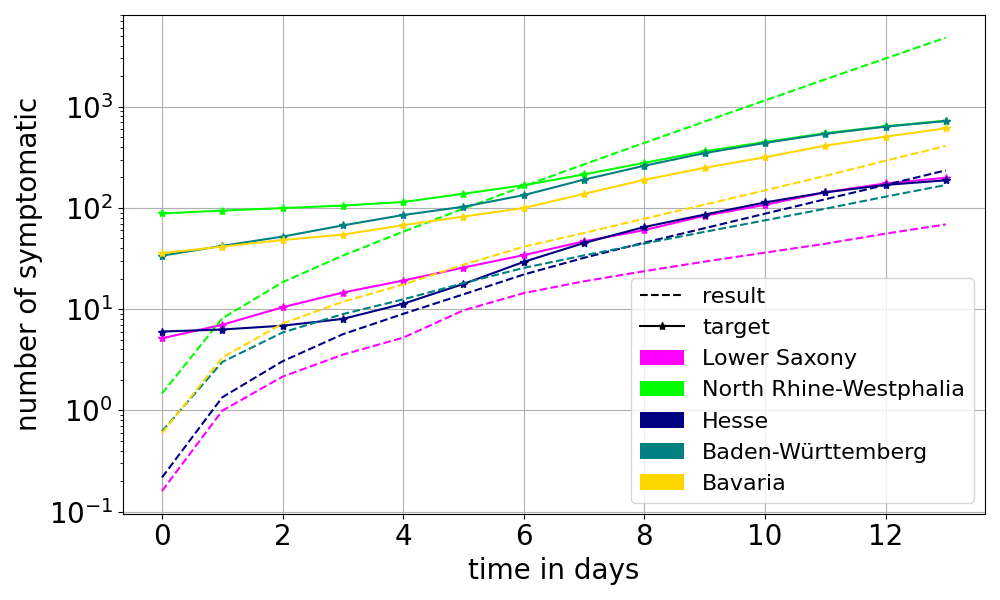}
        \caption{ODE federal states}
        \label{fig:result_ODE_without_NW}
    \end{subfigure}
    \caption{Results for the baseline scenario with total travel restrictions to and from North Rhine-Westphalia across the federal states of Germany, averaged over 10 simulation runs: simulated number of symptomatic individuals compared with the daily target data.}
    \label{fig:result_without_NW}
\end{figure}

The imposed travel restrictions reduce the number of symptomatic cases in Berlin. Due to the coupling of the federal states through mobility and subsequent transmission chains, the restrictions also affect the epidemic dynamics in all other federal states. Consequently, the average number of symptomatic cases decreases in some domains, while it increases in others. 
Three federal states exhibit an increased average number of symptomatic individuals when travel restrictions are imposed on Berlin: Mecklenburg-Vorpommern, Saxony, and North Rhine-Westphalia. 
Aside from Berlin, the largest reduction in the total number of symptomatic individuals is observed in Bavaria, where the number decreases from approximately $507$ to $358$. This indicates that the impact of travel restrictions is not solely determined by geographical proximity. 
Different choices of the restricted federal state or different initial distributions of infected individuals may therefore lead to qualitatively different outcomes, including an increase in symptomatic cases within the restricted state itself. 

North Rhine-Westphalia exhibits by far the largest number of incoming and outgoing jumps between federal states in our mobility data, making it a particularly interesting candidate for investigating the effects of regional border closures. As an example, we apply a complete border closure to North Rhine-Westphalia, which is represented as an ODE domain. The results, averaged over 10 simulation runs, are shown in Fig~\ref{fig:result_without_NW}. Comparing the results during the final simulation days, we observe that the number of symptomatic individuals in North Rhine-Westphalia is higher under the travel restriction scenario than in the baseline scenario. A possible explanation is that, without travel restrictions, North Rhine-Westphalia may temporarily lose more individuals through outgoing travel than it gains through incoming travel. Since the federal state is modeled as an ODE domain, such changes affect the number of individuals participating in the local epidemic dynamics. In contrast, under a complete border closure, the total population number remains fixed. 
Looking at the mean number of symptomatic individuals in Table~\ref{tab:border_closure}, the nationwide total under the North Rhine-Westphalia closure scenario is higher than in the baseline scenario. However, this increase is driven entirely by North Rhine-Westphalia itself: when its own contribution is excluded, the remaining federal states show a lower mean number of symptomatic individuals under this closure scenario than in the baseline.

Note that the implemented border restrictions are symmetric. For a system consisting of two connected domains, restricting travel between them yields the same mobility reduction regardless of whether the restriction is formally assigned to the first domain, the second domain, or both. 

\subsection{Zero-COVID and No-COVID} \label{subsec:zero_no_covid}

There are only limited details available on the practical implementation of the Zero-COVID and No-COVID strategies. Therefore, we used the paper~\cite{BA2023Reflections} and additional sources~\cite{aerzteblatt_no_covid,wikipedia_zero_covid} as guidance for our implementation. 

We implemented the Zero-COVID policy by initially setting it to be inactive in every federal state. At the end of each day, we evaluate the 7-day incidence by computing the sum of newly symptomatic cases over the previous seven days. 
Initially, we assume a 7-day incidence of zero. For the ABM domains, this information can be stored and tracked directly. We assign newly symptomatic individuals to their home facility, as this information is readily available and better reflects how positive cases are recorded in practice.
For the non-ABM domains, we identify the term $\Delta t \gamma I$ as representing the newly symptomatic cases in both models. In the ODE system, these values can be stored directly. In the PDE system, however, the integral over the entire PDE domain must be computed to obtain the total number of newly symptomatic individuals $\Delta t \gamma I$. At the end of each day, we check whether the 7-day incidence is at least 1. If so, the Zero-COVID restrictions are applied to the entire federal state. Once the 7-day incidence falls below 1, the restrictions are lifted again. Since the non-ABM models represent population groups by continuous state variables rather than discrete individuals, very small non-zero symptomatic cases can occur. Therefore, we deliberately use a threshold of 1. 
Whenever the restrictions are active, all individuals in the ABM remain at home. For the non-ABM domains, we again use the activity reduction rate to model the impact of the restrictions on the infection dynamics.

A similar approach was implemented for the No-COVID policy. The main differences are that restrictions are lifted after 14 days and that we do not count all symptomatic individuals, but only symptomatic individuals who were infected by an asymptomatic infectious individual (infectious compartment $I$). We refer to these as mystery cases. In the ABM, this requires tracking every infection event individually and storing whether the infector belonged to compartment $I$. Individuals identified as mystery cases are counted only once they become symptomatic, since they first progress through the exposed state $E$, then the infectious state $I$, before reaching the symptomatic state $\sY$.

For the non-ABM models, individual infection chains cannot be reconstructed because ODE and PDE models describe aggregated population compartments. Therefore, mystery cases cannot be identified directly as in the ABM. Instead, we estimate the expected number of mystery cases from the infection dynamics. Since individuals from both infectious compartments are assumed to be equally infectious, the probability that a newly infected individual was infected by an individual from compartment $I$ is approximated by
\begin{align*}
    P_\text{mystery(t)} &= \frac{|I|}{|I| + |\sY|}.
\end{align*} 
We then multiply this probability by the number of newly symptomatic individuals to estimate the number of mystery cases. This quantity is used to determine whether restrictions should be imposed or lifted, based on the cumulative number of mystery cases over a period of 14 days. As in the Zero-COVID policy, the corresponding sum must exceed one for restrictions to be activated.
Because newly infected individuals first enter the exposed compartment $E$ and are counted as mystery cases only once they become symptomatic, the relevant probability would, in principle, have to be evaluated at the time of infection rather than at the time of symptom onset. In the ODE and PDE models, however, individual infection histories are not available. For simplicity, we therefore use the current value of $P_\text{mystery(t)}$ as an approximation to the corresponding value at the time of infection. 

\begin{figure}[h!] 
    \centering
    \begin{subfigure}[b]{0.33\textwidth}
       \includegraphics[width=\linewidth]{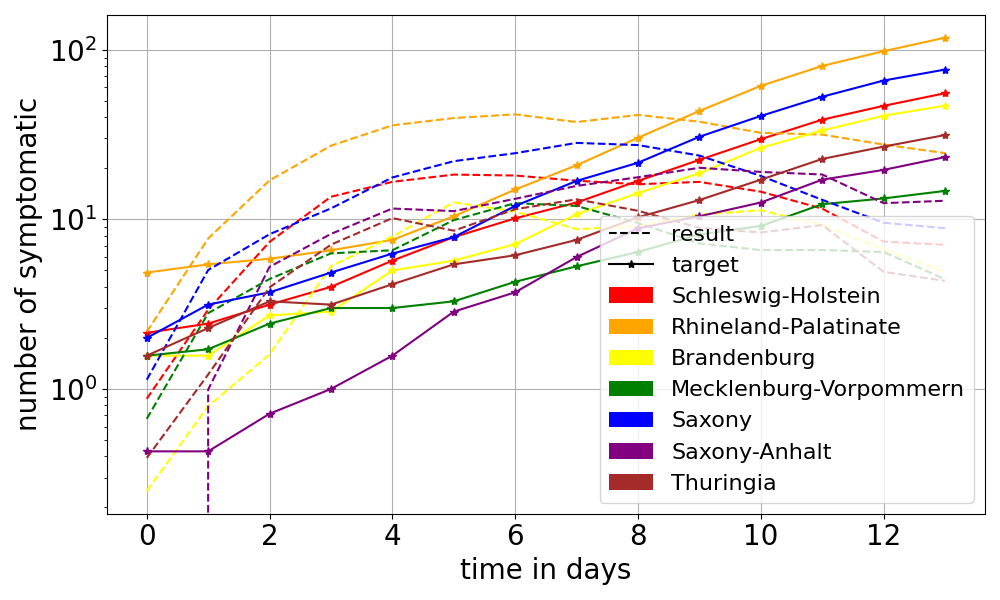}
        \caption{ABM federal states}
        \label{fig:result_zero_covid_ABM}
    \end{subfigure}
    \begin{subfigure}[b]{0.33\textwidth}
       \includegraphics[width=\linewidth]{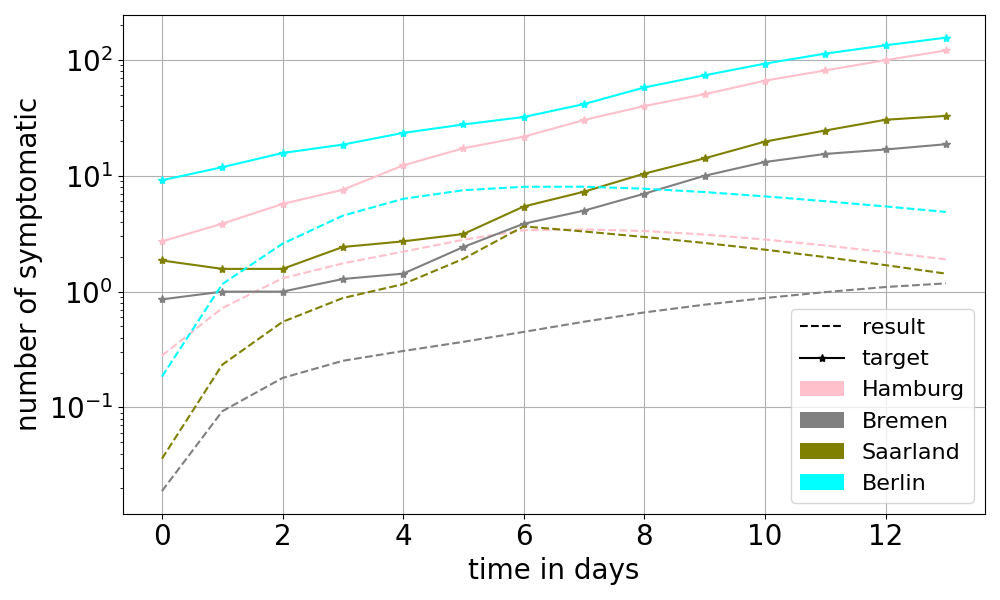}
        \caption{PDE federal states}
        \label{fig:result_zero_covid_PDE}
    \end{subfigure}
    \begin{subfigure}[b]{0.33\textwidth}
       \includegraphics[width=\linewidth]{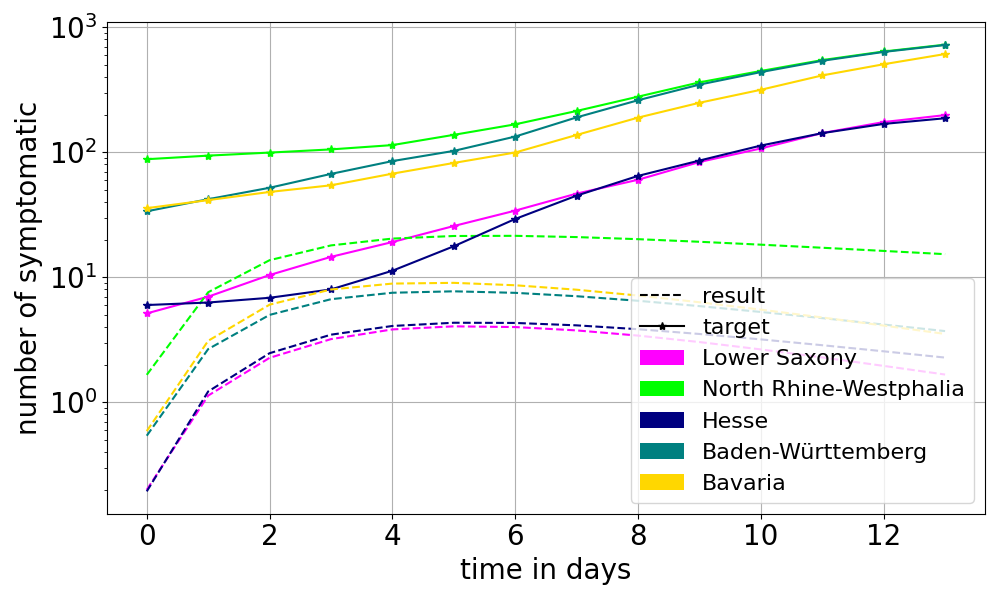}
        \caption{ODE federal states}
        \label{fig:result_zero_covid_ODE}
    \end{subfigure}
    \caption{Results for the Zero-COVID scenario across the federal states of Germany, averaged over 10 simulation runs: simulated number of symptomatic individuals compared with the daily target data.}
    \label{fig:zero_covid}
\end{figure}

\begin{figure}[h!] 
    \centering
    \begin{subfigure}[b]{0.33\textwidth}
       \includegraphics[width=\linewidth]{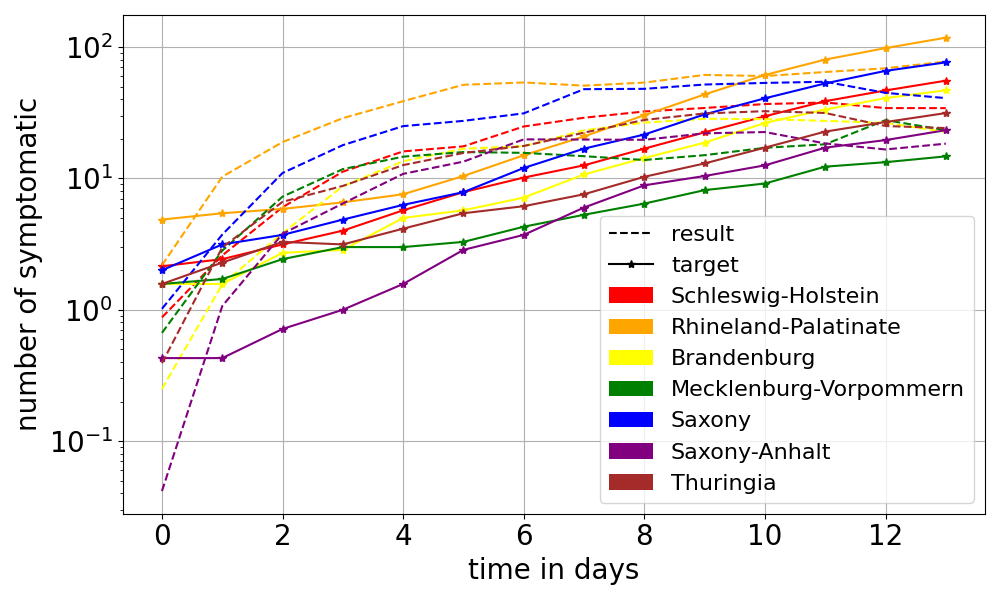}
        \caption{ABM federal states}
        \label{fig:result_no_covid_ABM}
    \end{subfigure}
    \begin{subfigure}[b]{0.33\textwidth}
       \includegraphics[width=\linewidth]{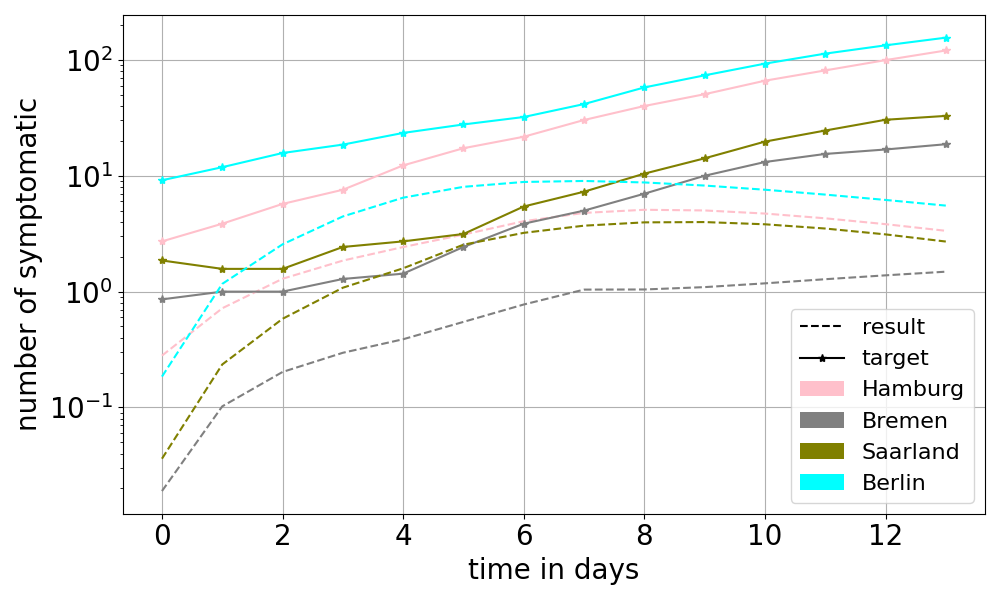}
        \caption{PDE federal states}
        \label{fig:result_no_covid_PDE}
    \end{subfigure}
    \begin{subfigure}[b]{0.33\textwidth}
       \includegraphics[width=\linewidth]{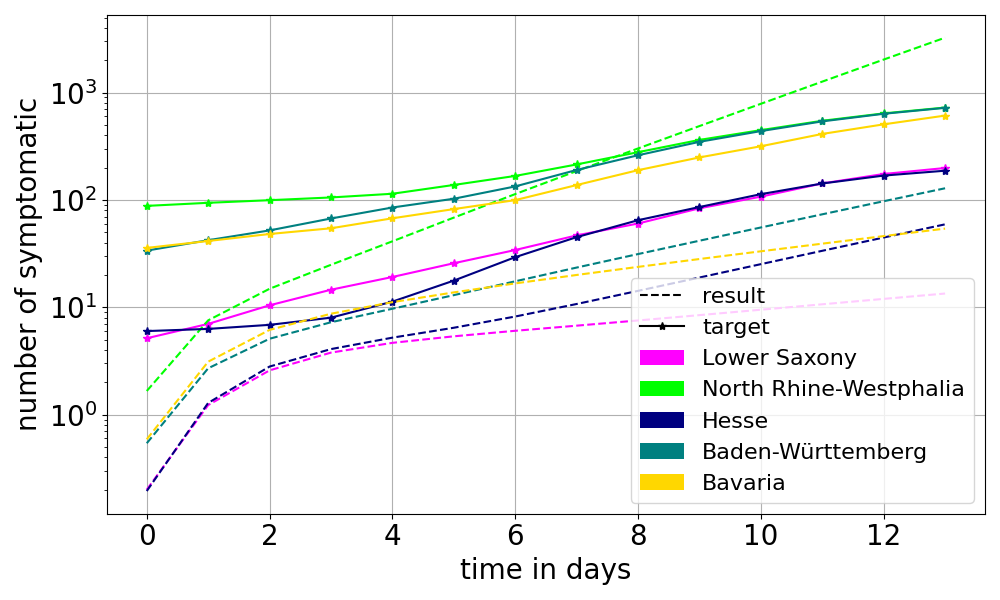}
        \caption{ODE federal states}
        \label{fig:result_no_covid_ODE}
    \end{subfigure}
    \caption{Results for the No-COVID scenario across the federal states of Germany, averaged over 10 simulation runs: simulated number of symptomatic individuals compared with the daily target data.}
    \label{fig:no_covid}
\end{figure}

The results are shown in Figs~\ref{fig:zero_covid} and~\ref{fig:no_covid}. Compared to the baseline scenario (see Fig~\ref{fig:result_Berlin_as_PDE}), the No-COVID scenario yields fewer symptomatic individuals on day 14 in all federal states except North Rhine-Westphalia, Schleswig-Holstein, Rhineland-Palatinate and Mecklenburg-Vorpommern. A comparison of the No-COVID and Zero-COVID scenarios shows that the Zero-COVID policy results in fewer symptomatic individuals on day 14 in every federal state. Table~\ref{table:no_zero_covid} provides a different perspective by reporting the mean number of symptomatic individuals in Germany averaged over the entire simulation period. The No-COVID policy reduces this average by 303 individuals compared to the baseline scenario, while the Zero-COVID policy yields a further reduction of 1042 individuals compared to the baseline scenario. Within the considered simulation period, the Zero-COVID policy therefore appears substantially more effective than the No-COVID policy. 

\begin{table}[!ht]
    \centering
    \begin{tabular}{c||c|c|c}
        scenario & baseline & No-COVID & Zero-COVID \\
        \hline
        mean number of symptomatic individuals  & $1175$ 
        & $872$ 
        & $133$
        \\
        absolute difference to baseline scenario & $0$ & $303$ & $1042$ \\
        mean number of symptomatic individuals without NW & $752$ & $259$ & $117$ \\
        absolute difference to baseline scenario without NW & $0$ & $493$ & $635$
    \end{tabular}
    \caption{Comparison of the baseline, No-COVID, and Zero-COVID scenarios. Mean number of symptomatic individuals in Germany, averaged over 10 simulation runs and the 14-day simulation period. Values are rounded to the nearest integer. NW denotes North Rhine-Westphalia.} 
    \label{table:no_zero_covid}
\end{table}

The results indicate that the No-COVID policy is more effective at reducing the number of symptomatic individuals in Germany than closing the borders of all federal states (see Tables~\ref{tab:border_closure} and~\ref{table:no_zero_covid}). However, this is no longer the case when North Rhine-Westphalia is excluded from the overall numbers. In that case, No-COVID performs worse than closing the borders of all federal states.
This is an interesting observation, because once the restrictions become active, individuals in the affected federal states are required to stay at home, which is considerably more restrictive than border closures. Nevertheless, the two interventions remain difficult to compare directly, since the border-closure scenario affects all federal states, whereas No-COVID may affect only a subset of federal states and not necessarily throughout the entire simulation period.

\noindent

\section{Discussion}

The computational time saved by simulating Berlin as a non-ABM domain rather than as an ABM domain may appear modest for a simulation period of only two weeks (see Table~\ref{table:times}). Assuming a similar effect for all federal states, the estimated simulation time for Germany modeled entirely with ABM domains would be approximately 4 hours for a two-week simulation, corresponding to about 103 hours for a full simulation year. In our previous work~\cite{Kehrer2026HybridAbmPde}, we evaluated relative changes between consecutive cumulative runs and showed that, depending on the chosen threshold, the Berlin–Brandenburg domain modeled entirely with ABMs required 92 simulation runs. Assuming the same number of runs is needed here, the total computational time without parallelization would amount to approximately 395 days. Since deterministic model components reduce the number of simulation runs required to achieve a given accuracy~\cite{Kehrer2026HybridAbmPde}, our hybrid ABM-PDE-ODE model in the baseline scenario would require only about 63 days to simulate an entire year with a sufficient number of runs. Although the difference may appear moderate at first glance, the reduction in computational effort is substantial, particularly for longer simulation periods. 

Although the infection rates could be further optimized, the parameter set used for the hybrid ABM-PDE model in our previous work provides a reasonable starting point for simulations involving ABM and PDE domains. If a spatial domain is sufficiently homogeneous, using similar infection rates for the PDE and ODE models appears reasonable based on the structure of the underlying equations. This observation is supported by our experiments: when Berlin was represented by an ODE model instead of a PDE model, the same correction terms could be retained while still producing comparable results. Similar behaviour was observed for Hamburg, which is likewise a relatively homogeneous federal state.
In contrast, the ODE federal states in the baseline scenario exhibit considerably stronger spatial heterogeneity. Consequently, the infection rates required adjustment when they were represented by an ODE model.

The Berlin border-closure experiment demonstrates that restricting travel to and from a single federal state can affect the epidemic dynamics throughout Germany, with the magnitude of these effects depending on the underlying mobility patterns and the current epidemic situation. Consequently, restricting travel to and from a federal state with high commuter traffic does not necessarily reduce the number of symptomatic individuals either within the restricted federal state or in the remaining federal states and may even produce the opposite effect in some regions. The presented results are based on a limited observation period and involve stochastic processes such as disease transmission and activity reductions. Therefore, a larger number of simulation runs would be required to obtain more robust conclusions. The experiment demonstrates that the consequences of mobility restrictions can be non-trivial and difficult to predict intuitively. Although the considered scenarios represent simplified abstractions of real-world processes, they provide useful insights into the potential consequences of mobility interventions. 

While the increase in symptomatic individuals after border closure is easiest to investigate for ODE federal states, it is difficult to generalize this observation to other scenarios. Nevertheless, it is useful to examine possible mechanisms. In additional experiments, border closures were applied separately to each ODE federal state. Among all considered ODE federal states, an increase in symptomatic individuals following border closure was observed only for North Rhine-Westphalia. A simple comparison of the total numbers of incoming and outgoing travelers does not appear sufficient to explain the observed effect. 
In the ODE model, the number of newly infected individuals is determined by the term $\frac{\beta \pi r^2}{|\Omega_\text{ODE}|} S(I+\sY)$. 
Since the susceptible population remains close to the total population size, variations in the number of infectious individuals have the strongest influence on the infection dynamics. Border closures modify mobility patterns and therefore the numbers of susceptible and infectious individuals present in a federal state. Small changes in the number of infectious individuals may consequently lead to noticeable differences in the cumulative number of infections. 
We note that North Rhine-Westphalia not only has the largest population among all federal states, but also at least 2.6 times as many initially exposed individuals. Furthermore, among the ODE federal states, it has the second-smallest area after Hesse. Based on these characteristics alone, one may expect stronger infection dynamics in North Rhine-Westphalia than in the other ODE federal states when borders are closed. 
Once mobility dynamics and stochastic health-state transitions in the ABM domains are taken into account, however, the evolution of compartment sizes becomes considerably more difficult to predict. Assuming that North Rhine-Westphalia exhibits the strongest infection dynamics in Germany, the proportion of infectious individuals among incoming travelers may be lower than among the local population. Under this assumption, allowing travel would redistribute infectious individuals from North Rhine-Westphalia to other federal states, while simultaneously replacing them with travelers who are less likely to be infectious. At the same time, visitors entering North Rhine-Westphalia would be exposed to stronger infection dynamics, whereas individuals from North Rhine-Westphalia traveling to other federal states would enter regions with weaker infection dynamics. Border closures would prevent this redistribution and retain a larger fraction of infectious individuals within North Rhine-Westphalia. This could contribute to higher infection numbers in North Rhine-Westphalia and lower infection numbers in the remaining federal states. 
Note that for the baseline scenario, the infection rates used for the ODE domains are larger than those used for the PDE domains. This may further contribute to stronger infection dynamics in North Rhine-Westphalia. 
However, the interaction between travel patterns, disease dynamics, and intervention measures is highly complex, making it difficult to predict the effects of travel restrictions from mobility statistics alone. 

We have shown that closing the borders of Berlin is less effective at reducing the number of symptomatic individuals in Germany than the No-COVID strategy. Nevertheless, a considerable number of symptomatic individuals remains under the No-COVID strategy during the considered simulation period. Since the number of symptomatic individuals is already decreasing in several federal states towards the end of the 14-day period, the long-term impact of the interventions may be larger than observed here. Consequently, longer simulation periods would be required to assess their full effect.
Furthermore, the most restrictive interventions are not necessarily the most effective in the long term, as they may delay infections~\cite{OECD2020Flattening}. 
Economic and societal costs~\cite{BoerschSupan2023Social,Podolsky2022Systematic} are also not considered in the present study, and perfect compliance with the imposed restrictions cannot be expected in reality. The real-world impact of such interventions may therefore differ from the simulated outcomes.
It is important to note that the restrictions associated with the Zero-COVID and No-COVID policies are identical in our simulation. The difference lies solely in the criteria used to activate and deactivate them. In both cases, the corresponding thresholds are evaluated once per day using cumulative counts over the previous 7 or 14 days, respectively.
Interestingly, not all federal states are subject to restrictions immediately or throughout the entire simulation period. Some federal states experience no restrictions at all in certain simulation runs, such as Rhineland-Palatinate in the No-COVID scenario and Bremen in the Zero-COVID scenario. Moreover, some states experience a lifting of restrictions despite the relatively short simulation period of 14 days. An example is Saarland in the Zero-COVID scenario.
Finally, restrictions may already be activated at an early stage of the epidemic, in some cases as early as March 3, 2020. Different starting conditions or later phases of the epidemic would likely lead to different effects and restriction durations.

Although 
we now explicitly store the IDs of agents transitioning between states, one issue remains unresolved. In some cases, agents enter a PDE or ODE state in an infected health state and subsequently leave it as susceptible. This behavior is inconsistent with the epidemiological model. Naive implementations of the additional storage of travelers' health states and the computation of their probable health states based on visit duration can lead to non-negligible increases in both computational time and memory consumption. While the mobility model considered in~\cite{zunker2026efficient} differs from the approach used in this work, the authors address a similar problem, namely the efficient computation of traveler states, and propose numerical methods to reduce the associated computational costs. 

\section{Conclusion} 

In this work, we presented a multiscale, mobility-informed framework that couples agent-based models (ABMs), partial differential equation (PDE) models, and ordinary differential equation (ODE) models within a unified epidemic simulation environment. The proposed approach enables different regions to be represented at different levels of resolution while preserving interactions between them through mobility-driven transitions, restricting computationally expensive agent-based simulations to selected regions of interest while using aggregated PDE and ODE descriptions elsewhere.

The numerical experiments demonstrated that the framework can reproduce realistic epidemic dynamics on a national scale while maintaining flexibility in the choice of model resolution. Comparisons with reported infection data showed that the coupled ABM-PDE-ODE model is capable of capturing the overall development of the epidemic despite the use of different model resolutions across federal states. The results further indicate that sufficiently homogeneous regions can be exchanged between ABM, PDE, and ODE descriptions without fundamentally altering the qualitative behaviour of the overall system, suggesting that high-resolution modeling is not necessarily required everywhere. 


The intervention experiments highlighted the importance of mobility-driven interactions between regions. In some cases, regional border closures increased infections within the restricted region and contributed to higher infection numbers at the national level. This suggests that the effect of intervention depends strongly on how regions are connected and how their epidemic dynamics differ. 

Beyond the specific COVID-19 case study, the proposed framework provides a general methodology for coupling individual-based and population-based epidemic models that is not tied to a particular disease, mobility dataset, or geographical setting. It is therefore well suited for investigating how local epidemic dynamics interact with large-scale mobility networks and how model resolution influences simulation outcomes more broadly.

Several directions for future work remain. The present study focused on a limited simulation period and a specific set of intervention strategies. Future investigations could consider longer time horizons, additional intervention measures, and systematic analyses of mobility networks to identify groups of strongly connected regions. Such analyses may help determine which regions require high-resolution modeling and whether these groups provide a more suitable basis for epidemic interventions than federal state boundaries. This idea is consistent with approaches based on mobility functional areas, where highly connected regions are identified from observed human mobility patterns rather than predefined administrative boundaries~\cite{Iacus2022Mobility}. 
Furthermore, the framework could be extended to other infectious diseases, alternative mobility datasets, larger geographical regions, or additional modeling approaches.

\section*{Acknowledgments} 
We acknowledge Kai Nagel for valuable discussions on the agent-based model. We also thank Jakob Benjamin Rehmann and Sydney Cornelia Paltra for providing essential data for the Germany experiments and for assistance with their software. 
We would like to thank Inan Bostanci for the many meetings and constructive feedback throughout the process.



This work was funded by the German Federal Ministry of Research, Technology and Space (BMFTR) through the project EPISERVE (project ID: 031L0324A) and through the project MODAL (fund numbers 05M2025). EPISERVE is part of the German Modeling network for severe infectious diseases (MONID). Further funding was provided by the Deutsche Forschungsgemeinschaft (DFG, German Research Foundation) under Germany´s Excellence Strategy – The Berlin Mathematics Research Center MATH+ (EXC-2046/1, EXC-2046/2, project ID: 390685689).
\clearpage

\bibliography{references}{}
\bibliographystyle{vancouver}




%
\end{document}